\documentclass[twocolumn]{aastex631}
\shorttitle{\(\eta\) vary with \(\Sigma\)}
\shortauthors{Zhong et al.}
\graphicspath{{./}{figures/}}
\usepackage{times}
\usepackage{color}
\accepted{September 26, 2023}

\begin{document}
	
\title{Kinematical fluctuations vary with galaxy surface mass density}

\author[0009-0009-0648-0993]{Ze-Hao Zhong}
\affiliation{CAS Key Laboratory of Optical Astronomy, National Astronomical Observatories, Chinese Academy of Sciences, Beijing, 100101, China}
\affiliation{School of Astronomy and Space Science, University of Chinese Academy of Sciences, Beijing, 100049, China}
\affiliation{Max-Planck-Institut f\(\ddot{u}\)r Astronomie, K\(\ddot{o}\)nigstuhl 17, D-69117 Heidelberg, Germany}

\author[0000-0002-8980-945X]{Gang Zhao}	
\affiliation{CAS Key Laboratory of Optical Astronomy, National Astronomical Observatories, Chinese Academy of Sciences, Beijing, 100101, China}
\affiliation{School of Astronomy and Space Science, University of Chinese Academy of Sciences, Beijing, 100049, China}

\author[0000-0003-4996-9069]{Hans-Walter Rix}
\affiliation{Max-Planck-Institut f\(\ddot{u}\)r Astronomie, K\(\ddot{o}\)nigstuhl 17, D-69117 Heidelberg, Germany}

\author[0000-0001-6947-5846]{Luis C. Ho}
\affiliation{Kavli Institute for Astronomy and Astrophysics, Peking University, Beijing 100871, China}
\affiliation{Department of Astronomy, School of Physics, Peking University, Beijing 100871, China}

\correspondingauthor{Gang Zhao}   
\email{gzhao@nao.cas.cn}

\begin{abstract}
The Galaxy inner parts are generally considered to be optically symmetric, as well as kinematically symmetric for most massive early-type galaxies. {At the lower-mass end,} many galaxies contain lots of small patches in their velocity maps, causing their kinematics to be nonsmooth in small scales and far from symmetry. These small patches can easily be mistaken for measurement uncertainties and have not been well discussed. We used {the comparison} of observations and numerical simulations to demonstrate its existence beyond uncertainties. For the first time we have found that the fluctuation degrees have an approximate inverse loglinear relation with the galaxy stellar surface mass densities. This {tight relation among galaxies that do not show obvious optical asymmetry that traces environmental perturbations} indicates that stellar motion in galaxies has inherent asymmetry besides external environment influences. The {degree of the kinetic asymmetry} is closely related to and constrained by the intrinsic properties of the host galaxy.
\end{abstract}
\keywords{Galaxy kinematics (602); Galaxy dynamics (591); Galaxy structure(622); Dark matter(353)}

\section{Introduction} \label{sec1}
\begin{figure*}[!ht]
\gridline{\fig{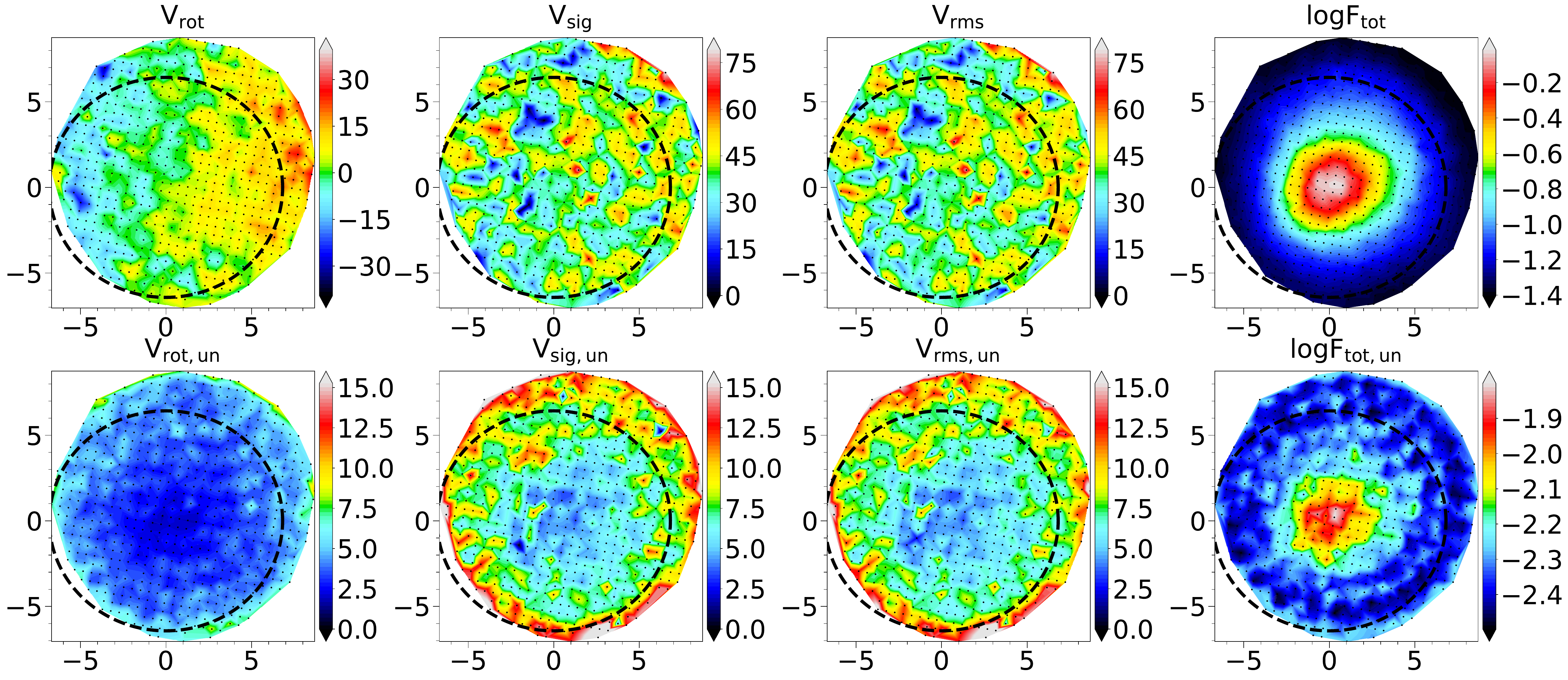}{0.45\textwidth}{}\fig{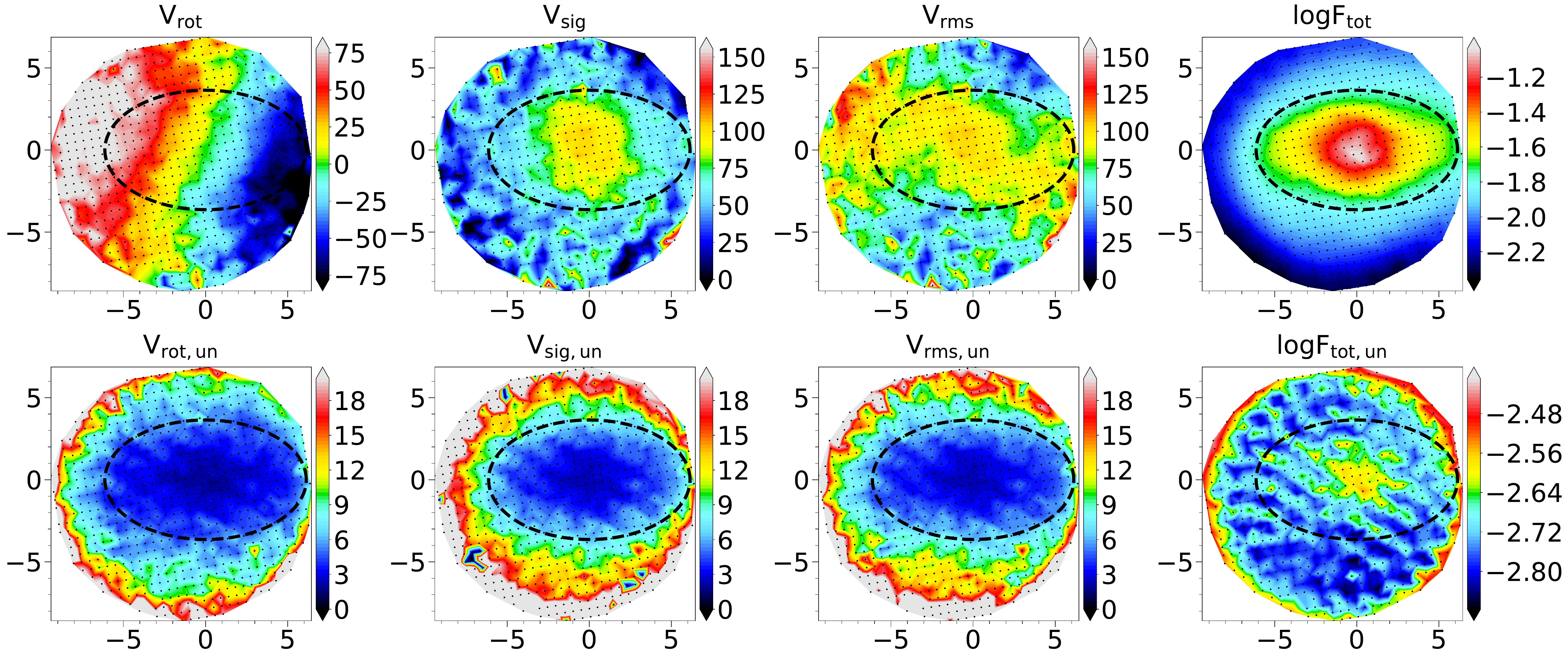}{0.45\textwidth}{}}
\caption{Comparison of small-scale fluctuation degrees among galaxies with different masses. Left panel: {The four upper maps represent velocity, velocity dispersion, total velocity, and flux distribution, respectively, and the four lower maps represent the corresponding uncertainty maps} of SAMI70114 within effective radius \(R_{\rm e}=0.685\ \rm kpc\). The stellar mass is \(\log(M/M_{\rm \odot})=9.03\) and redshift is \(z=0.0038\). The units of velocity are \(\rm km\,s^{-1}\), flux units are \(10^{-16}\ \rm erg\,s^{-1}\,cm^{-2}\,\mathring{A}^{-1}\,spaxels^{-1}\) and the coordinate axis is in arcseconds. Right panel: same as the left panel, but for SAMI56064, with \(R_{\rm e}=4.65\ \rm kpc\), \(\log(M/M_{\rm \odot})=10.44\) and \(z=0.0404\). The instrumental resolution of the SAMI survey is 70 \(\rm km\,s^{-1}\), but stellar velocity being lower than half instrumental resolution may also be reliable using the new spectrum fitting penalized pixel-fitting (PPXF) method \citep{2004PASP..116..138C,2017MNRAS.466..798C}. The two examples show that low-mass galaxies tend to have higher small-scale fluctuation degrees compared with massive galaxies. \label{fig1}}
\end{figure*}

Although there may be some lopsidedness in the outer parts of galaxies, their inner parts are generally considered to be optically symmetric \citep{1995ApJ...447...82R,1998AJ....116.1163R,2000ApJ...529..886C}, suggesting that their distribution of matter is relatively smooth. Therefore, widely used dynamical models, such as the spherical model, the axisymmetric model \citep{2008MNRAS.390...71C} or the Schwarzschild model \citep{1979ApJ...232..236S}, usually assume some symmetry in the distribution of matter. Under these gravitational potentials, stellar kinematics are generally centrosymmetric after being projected onto a two-dimensional plane. Massive early-type galaxies usually observed as symmetric stellar motions, which have provided good evidence for the above assumption \citep{2006MNRAS.366.1126C,2013MNRAS.432.1709C}. However, there are also some galaxies classified as nonregular rotations, which causes their stellar kinematics to be complex  \citep{2016ARA&A..54..597C}. {Observations} from some integral field
spectrograph (IFS) surveys have shown that the stellar velocity fields of many less massive galaxies do not {appear} to be centrosymmetric or smooth; this is also valid even for a few massive galaxies \citep{2013MNRAS.432.1709C,2019MNRAS.490.2124L}. {Moreover}, stars can have many noncircular motions, and it is difficult for the stellar motion in every galaxy to be {perfectly} symmetric. So what kind of galaxy is more kinematically symmetric? As shown in Figure \ref{fig1}, the velocity maps of {low-mass galaxies appear to} be more asymmetric compared to those of more massive galaxies. Nevertheless, due to the larger measurement uncertainties of low-mass galaxies, it is difficult to distinguish the real asymmetry from the measurement uncertainty of the velocity field. That is, the two are observationally degenerate. Thus there have also been few statistical studies of the asymmetry of stellar motions in low-mass galaxies until now. However, although accounting for the relatively larger measurement uncertainties of low-mass galaxies, we found that uncertainty is insufficient to fully explain these small patches. These patches are just like {physical fluctuations that are not caused by mesurement uncertainties} on a smooth and perfectly symmetric image, so we call them kinematical fluctuations, {with respect to fluctuation in surface brightness, which is referred to as "optical asymmetry."} {We note that, with} more small-scale fluctuations, the kinematics will be more asymmetry.

%\gridline{\fig{}{width=0.9\linewidth}{(a)}}
%\gridline{\fig{}{width=0.9\linewidth}{(b)}}

In this work, we aim to quantitatively describe the stellar kinematical asymmetry of low-mass galaxies and try to find the most related influence that causes it to be asymmetry. Then, we quantify their small-scale fluctuation degree and thus the {kinetic} asymmetry degree of the galaxies. Generally, morphology and kinematics are the two aspects involved in the study of galaxy symmetry. The methods for measuring luminous morphological symmetry usually adopt \(180^{\circ}\) rotation self-subtracting methods \citep{2000ApJ...529..886C,2003ApJS..147....1C,2004AJ....128..163L} or Fourier decomposition \citep{1995ApJ...447...82R,1998AJ....116.1163R,2010AJ....139.2097P,2017ApJS..232...21K}. In addition, {\small KINEMETRY} \citep{2006MNRAS.366..787K} is a widely used method and also applies Fourier decomposition to assess the stellar kinematical symmetry degree of massive galaxies in many IFS surveys (SAURON, \citealt{2008MNRAS.390...93K}; \(\rm ATLAS^{3D}\), \citealt{2013MNRAS.432.1768K}), or gas kinematical symmetry degree \citep{2008ApJ...682..231S,2010ApJ...724.1373G,2012MNRAS.424.2232A,2017MNRAS.465..123B,2020ApJ...892L..20F}. {\small KINEMETRY} is useful when studying the velocity and velocity dispersion for the asymmetry degree and nonsmoothness, respectively, {and this method can also be applied to measure the rotation of nonregular rotating galaxies}. However, this method relies on fitting; for galaxies with large measurement uncertainties such as dwarf galaxies, it is more difficult to interpret the results when Fourier decomposition is applied. Besides the uncertainties, when the intrinsic irregularity of the galaxy is large, it is also hard to explain the more complex details by discussing velocity and velocity dispersion separately. {In another aspect}, gas kinematical asymmetry {appears} to be weakly inversely related to stellar mass of galaxies \citep{2017MNRAS.465..123B}. The large scatter of their trend is reasonable because gas motion is affected by many nongravitational factors, such as turbulence and shocks, so the intrinsic dispersion of the gas asymmetry is in principle large among a diversity of galaxies. With such a large dispersion, it is difficult to find the most related influence of the asymmetry. Stellar motion is mainly dominated by gravity, and the influencing factors are much less in comparison, so studying stellar kinematical asymmetry is more suitable to finding the specific influence from this aspect. Moreover, the outer part of the galaxy is more affected by galaxy interactions or mergers, so the relatively large asymmetry degree in the galaxy outer part is {expected}. The measurement uncertainties of the outer part are also larger owing to the decrease in brightness, so we here only focus on the inner part, thus within the effective radius \(R_{\rm e}\) of the galaxy. Therefore, We employ a new nonparametric method that is similar to the optical \(180^{\circ}\) rotation self-subtracting method, but we use it to study the stellar, total kinematics asymmetry within \(R_{\rm e}\) in galaxies. {In addition, in this paper "low-mass" refers to galaxies of \(\log(M/M_{\rm \odot})\leq 9.5\), which is approximately the largest stellar mass of dwarf galaxies. “Small scale” refers to the scale of \(\rm \sim 0.1\ kpc\) or several times \(\rm 0.1\ kpc\), and “large scale” means the scale of galaxy size or the size of its effective radius.}

We introduce our quantitative method in Section \ref{sec2}. In Section \ref{sec3}, we present the newly found inverse loglinear relation between kinematical fluctuation degree and galaxy stellar surface mass density, demonstrated by the {comparison} of observation and simulation data. We discuss the influence of measurement uncertainty and summarize our results in Section \ref{sec4}.

\section{METHODS}\label{sec2}
We choose to study the stellar rms velocity (\(V_{\rm rms}\)) maps obtained from Data Release 3 (DR3) of the SAMI Galaxy Survey \citep{2015MNRAS.447.2857B,2016MNRAS.463..170C,2017MNRAS.468.1824O,2017ApJ...835..104V,2021MNRAS.505..991C,2021MNRAS.504.5098D}, because SAMI is one of the IFS surveys, which are generally spatially resolved with local values, and SAMI contains a lower mass range of galaxies compared with other IFS surveys. Here \(V_{\rm rms}=\sqrt{(V_{\rm rot}^2+V_{\rm sig}^2)}\) for every velocity field {spatial pixel (spaxel)}, \(V_{\rm rot}\) is the local velocity, and \(V_{\rm sig}\) is the velocity dispersion {both obtained from spectrum fitting of the spaxel}. So \(V_{\rm rms}\) can show the total amount of "velocity," which also represents the line-of-sight stellar kinetic energy per unit mass.

Our new nonparametric approach to quantitatively study stellar asymmetry degree does not require any assumptions or fits to the data, just \(180^{\circ}\) rotation self-subtracting the squared value of \(V_{\rm rms}\) maps. {Before the rotation, we do not change the orientation and inclination of the galaxies in the intergral field unit (IFU) data cube, that is, the orientation of the galaxies is arbitrary as observed. Then, the \(180^{\circ}\) rotation is performed relative to the optical center of the galaxy and the coordinate axis of the IFU data cube. All the \(180^{\circ}\) rotation self-subtracting for other data cubes are also performed in the same way in this article. Since the galaxy centers of our sample are all at the intersection of the surrounding four spaxels, the \(180^{\circ}\) rotation can be performed conveniently.} We then use the ratio of \(180^{\circ}\) differences and the original mean value to define the asymmetry parameter \(\eta\). The corresponding {equation} is as follows:
\begin{equation}
	\eta_{i}=\frac{\left| V_{i}^2-V_{i,\,180}^2\right|}{(V_{i}^2+V_{i,\,180}^2)/2}
\end{equation}  
Here \(\eta_{i}\) is the asymmetric parameter of spaxel \(i\), and \(V_{i}\) and \(V_{i,\,180}\) are the \(V_{\rm rms}\) of the \(i\)th spaxel and its \(180^{\circ}\) rotated one, respectively. In order to study the kinematics of galaxy inner parts, we only calculate \(\eta_{i}\) within the effective radius \(R_{\rm e}\). For each galaxy, the total \(\eta\) is the light-weighted or mass-weighted average of all spaxels within \(R_{\rm e}\). It can be imagined that if there are more irregular small-scale fluctuations, the value of the asymmetry parameter \(\eta\) will be larger. It should be noted that if the IFU spaxels do not fill the effective radius (e.g., the left panel of Figure \ref{fig1}), then the IFU spaxels at the corresponding symmetric positions will also be truncated when calculating the total \(\eta\) (discard a few spaxels of the left panel of Figure \ref{fig1} at the corresponding right positions), so that all contained spaxels satisfy a symmetric distribution.

If galaxies contain too few spaxels, the effect of measurement uncertainty can be large, so we only select galaxies with more than 50 spaxels within \(R_{\rm e}\). On the other hand, for low-mass galaxies with relative larger uncertainties, {a} few \(V_{\rm rms}\) values {can be} greater than the escape velocity, which are less reliable. {We here use \(V_{\rm escape}=\sqrt{GM_{\rm dyn}/R_{\rm e}}\) to calculate the escape velocity because the total mass is approximately \(M_{\rm dyn}/2\) within \(R_{\rm e}\) assuming constant mass-to-light ratios, and \(M_{\rm dyn}\) is calculated by the virial estimator of \citet{2006MNRAS.366.1126C} (see Appendix C).} However, if the majority of spaxels are reliable, the small number of unreliable spaxels contributes less to the weighted average \(\eta\). Therefore, we use the escape velocity to classify the sample. Galaxies with all the spaxels being lower than the escape velocity within \(R_{\rm e}\) are called the pure sample, whose \(\eta\) values are relatively credible. The galaxies with spaxels exceeding the escape velocity are called the sub sample, whose value is used as a reference. {Most galaxies in the SAMI survey are not showing features associated with interaction \citep{2018MNRAS.476.2339B}. The main observational catalog CubeObs of SAMI DR3 offers a flag to excluded galaxies with irregular stellar kinematics due to nearby objects or mergers by visual inspection \citep{2017ApJ...835..104V,2021MNRAS.505..991C}. Here we also adopt this flag to exclude these galaxies; thus, our sample galaxies can be regarded as not in the phase of interaction or merger.}

\section{RESULTS}\label{sec3}
\begin{figure*}[!ht]
	\gridline{\fig{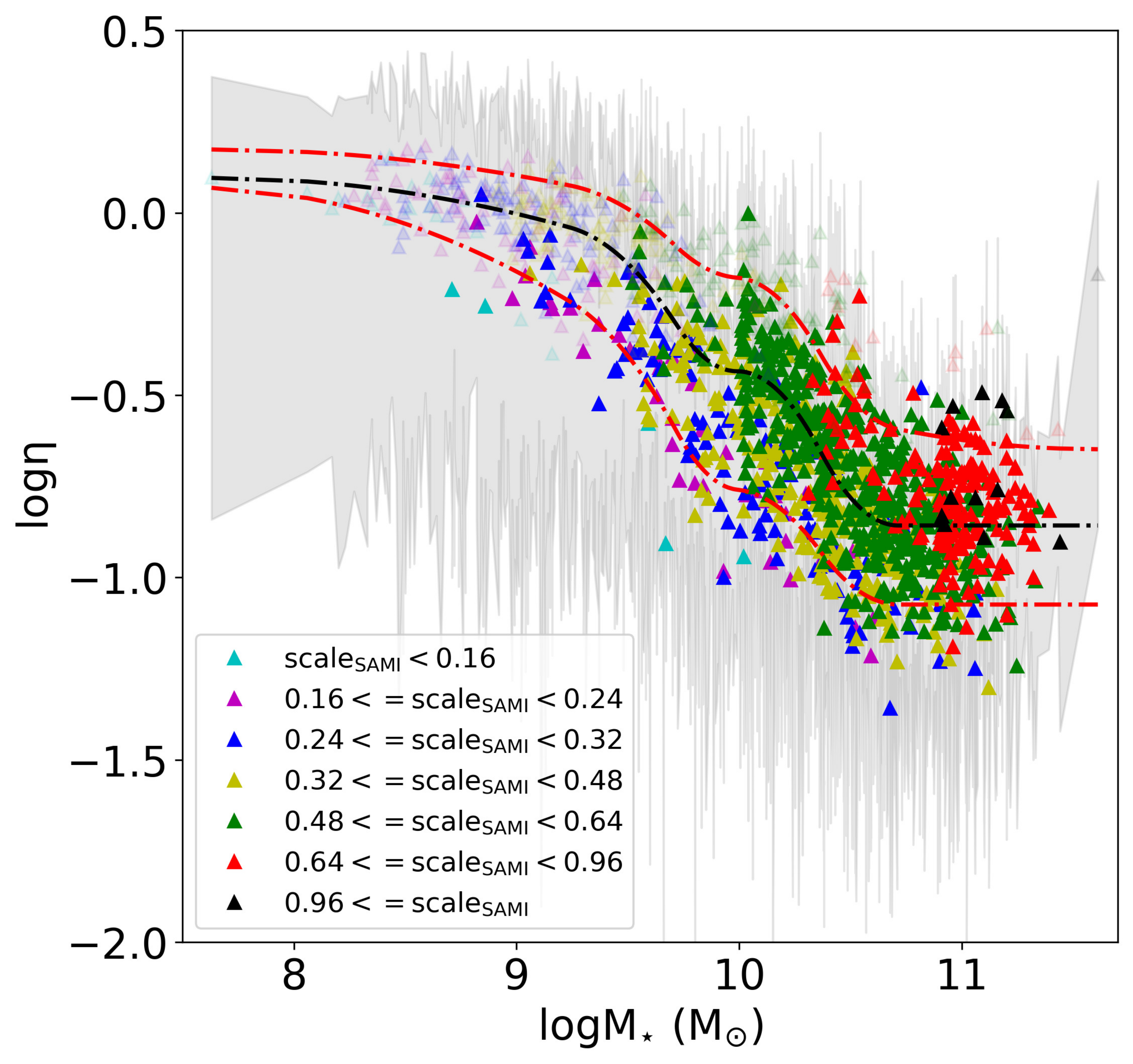}{0.45\textwidth}{}\fig{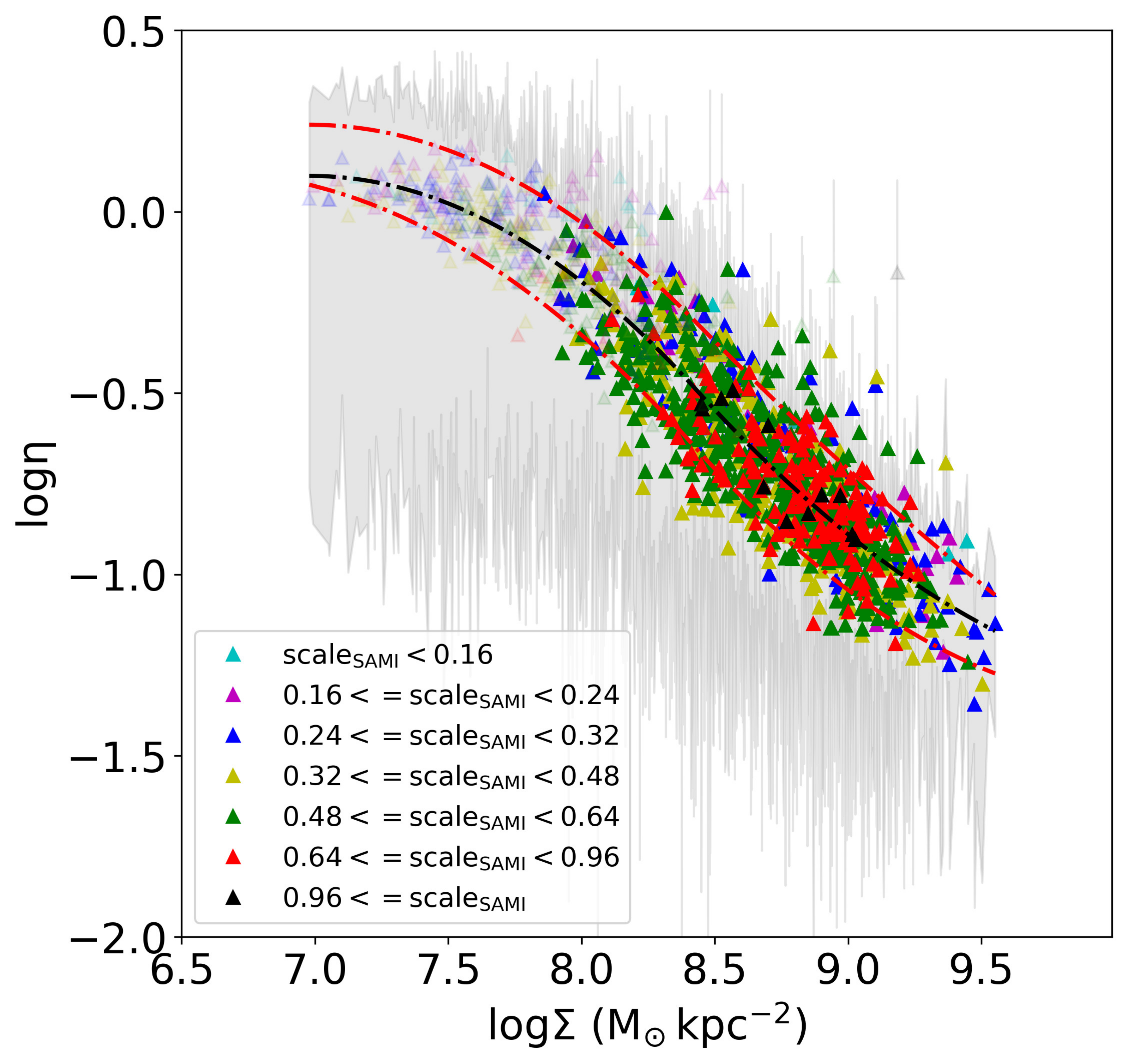}{0.45\textwidth}{}}
	\caption{\(\eta\) relations of SAMI galaxies. Left panel: relation of \(\eta\) with galaxy stellar mass. The colors represent different scale groups, and the transparent triangles represent subsamples. The gray shading is calculated from the maximal contribution of \(V_{\rm rms}\) uncertainties, and thus the maximal uncertainties of \(\eta\). The units for all scales are kpc. {The upper red, middle black, and lower red lines are 10\%, 50\%, and 90\% quantiles respectively.} Right panel: relation of \(\eta\) with galaxy stellar surface mass density within \(R_{\rm e}\). \label{fig2}}
\end{figure*}

We found that the asymmetry parameter \(\eta\) is closely related to the stellar mass and even more tightly related to the stellar surface mass density of galaxies of all masses, {as is shown in Figure \ref{fig2}}. {In order to show the trend of the relationship between parameters and the corresponding relation scatter, we applied the COnstrained B-Spline (cobs) quantile regression method in the R programming language \citep{cobs2007,cobs2022}. The dashed-dotted lines represent 90\% (upper red), 50\% (black middle), 10\% (lower red) quantile lines respectively in Figure \ref{fig2}. The red quantile line intervals can indicate that there is small relation scatter for both panels.} The relation for stellar mass is shown in the left panel of Figure \ref{fig2}; from \(\log(M/M_{\rm \odot}) \sim 9\) to \(\log(M/M_{\rm \odot}) \sim 10.5\) or so, the larger the stellar mass, the smaller the \(\eta\). \(\log(M/M_{\rm \odot}) \sim 10.5\) itself is also a well-known knee point of galaxies, {which related to the extreme point of baryon mass fraction in galaxies \citep{2010MNRAS.404.1111G,2013MNRAS.432.1709C,2018MNRAS.481.1950L}, or the disappearance of the trend of gas rotation with gas mass fraction \citep{2012ApJ...756..113H}}. {We considered that the trend in the left panel of Figure \ref{fig2} is real, and the small-scale fluctuations may also be related to the mass distribution in galaxies.} Furthermore, \(\eta\) have an approximate inverse loglinear relation with the stellar surface mass density of the galaxy (right panel of Figure \ref{fig2}), and this relation appears to be both tighter and better than that with stellar mass. The corresponding linear fitting slope is about \(-0.632\), and the gray shading in figures represents the maximal uncertainty contributions; {both will be discussed in detail later}. The different galaxy distances will lead to different IFU spaxel scales. We then artificially divided these galaxies into seven groups based on their distance from us, also corresponding to different scales, as shown in Figure \ref{fig2} (details given in Appendix A). {On the one hand, galaxies of different distances lie in the same narrow relation of the right panel of Figure \ref{fig2}. {This indicates that the relation is robust, independent of physical scales of IFU spaxels.} On the other hand, an IFS survey of the nearby universe usually cannot completely avoid observational selection effects (also displayed in Appendix A). The left panel of Figure \ref{fig2} is obviously affected by the selection effect; the trend scatter can be overestimated or underestimated by the observed sample number of corresponding mass, but the right panel is less influenced. This is also the reason we considered the relationship in the right panel of Figure \ref{fig2} to be better.}

Due to the huge diversity of galaxies, the correlations between different galaxy parameters usually have large intrinsic scatter, so tighter relationship between parameters often reveals their more fundamental physical connections. The tightly inverse loglinear relation of the right panel of Figure \ref{fig2} not only quantitatively describes {that} low-mass galaxies should be accompanied by higher stellar kinematical fluctuations but also {indicates} that the asymmetry of star motion inside galaxies is not {universal at all mass scales} but {dynamically} influenced by the galaxies' matter distribution.

\begin{figure*}[!ht]
	\gridline{\fig{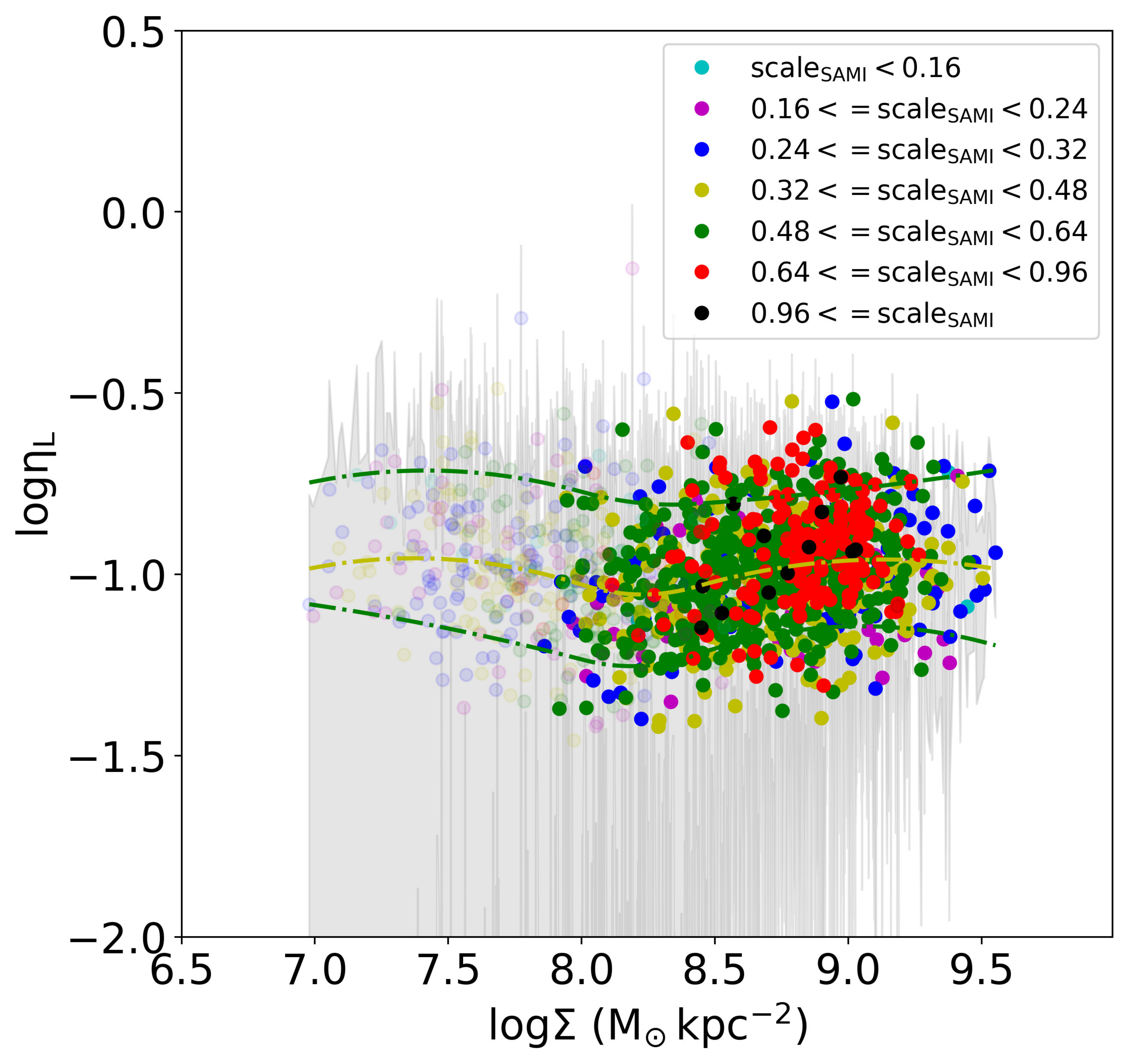}{0.45\textwidth}{}\fig{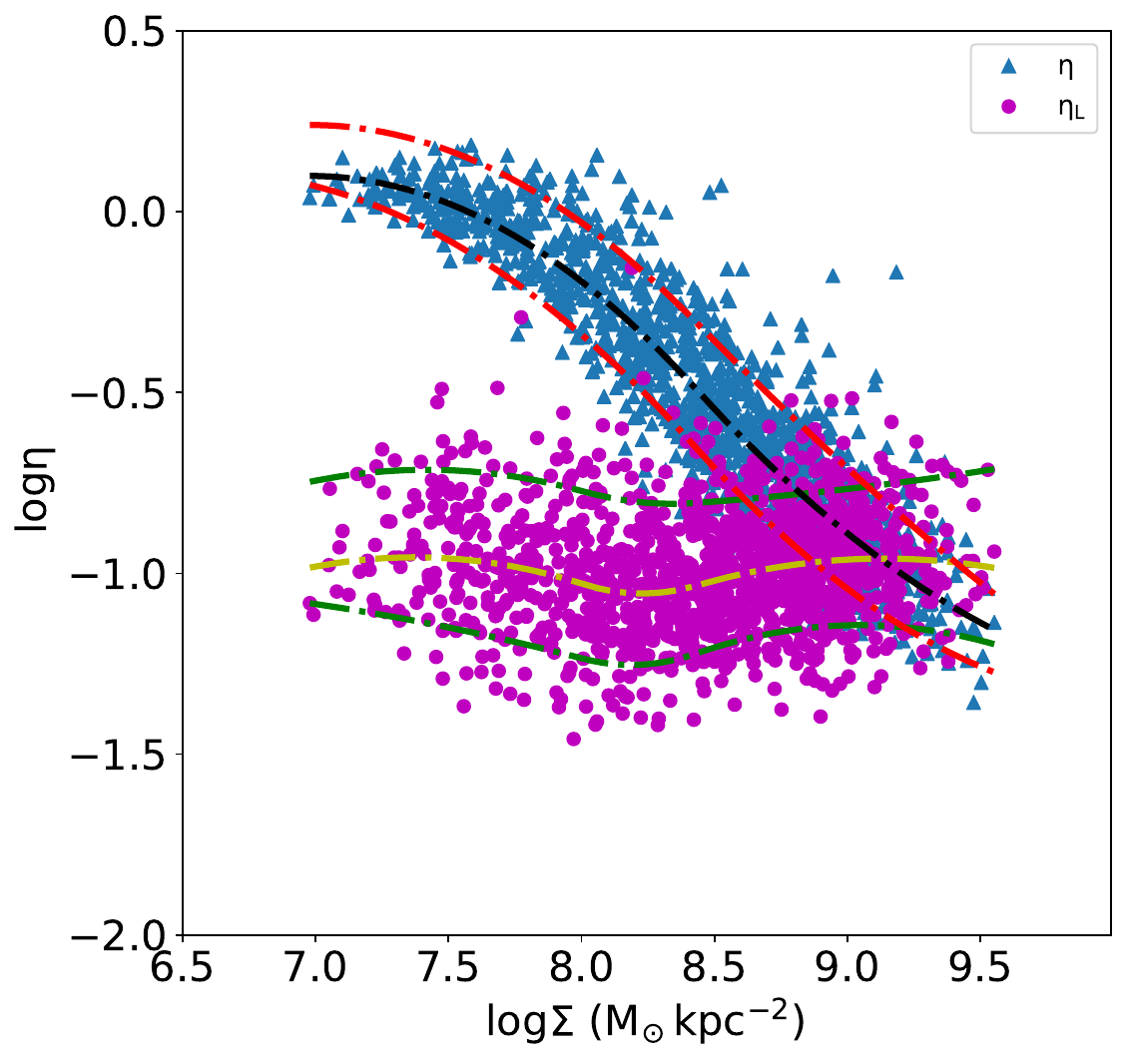}{0.45\textwidth}{}}
	\caption{{Optical asymmetry parameter \(\eta_{\,L}\) and its comparison with \(\eta\). Left panel: relation of \(\eta_{\,L}\) with galaxy stellar surface mass density within \(R_{\rm e}\). The colors, transparent circles and scale unit are the same as in Figure \ref{fig2}. The gray shading here is calculated from the maximal contribution of flux uncertainties, and thus the maximal uncertainties of \(\eta_{\,L}\). The upper green, middle yellow, and lower green lines are 10\%, 50\%, and 90\% quantiles, respectively. Right panel: comparison of optical and kinematical symmetries. In order to avoid visual confusion caused by colors of different sample groups, all \(\eta\) are represented by blue triangles, and all \(\eta_{\,L}\) are represented by magenta circles.} \label{fig3}}
\end{figure*}

{We note that these kinematically asymmetric galaxies are in fact optically symmetric within \(R_{\rm e}\). To demonstrate this, we} {calculated the optical asymmetry parameter \(\eta_{\,L}\), whose equation is similar to the asymmetry "A" used in previous works \citep{2000ApJ...529..886C,2003ApJS..147....1C,2004AJ....128..163L}. However, we have slightly adjusted the equation to keep it consistent with the kinematical asymmetry parameter \(\eta\). The calculation equation is as follows:
\begin{equation}
	\eta_{\,L,i}=\frac{\left| I_{i}-I_{i,180}\right|}{(I_{i}+I_{i,180})/2}
\end{equation} 
\(I_{i}\) and \(I_{i,180}\) are the flux of the \(i\)th spaxel and its \(180^{\circ}\) rotated one. Figure \ref{fig3} shows the distribution of \(\eta_{\,L}\) with galaxy stellar surface mass density and its comparison with \(\eta\). It can be seen from the figure that the optical asymmetry parameters \(\eta_{\,L}\) of all galaxies are similar, and the \(\log \eta_{\,L}\) of the 50th percentile line are all around \(-1.0\) for galaxies with any surface mass density. {This suggests that} they are all small-scale optically symmetrical. Although the values of \(\eta_{\,L}\) and \(\eta\) cannot be directly compared, the inverse loglinear relationship for \(\eta\) does not exist for \(\eta_{\,L}\), indicating that low-mass galaxies are optically symmetric but kinematically asymmetric.
}

\begin{figure*}[!ht]
	\gridline{\fig{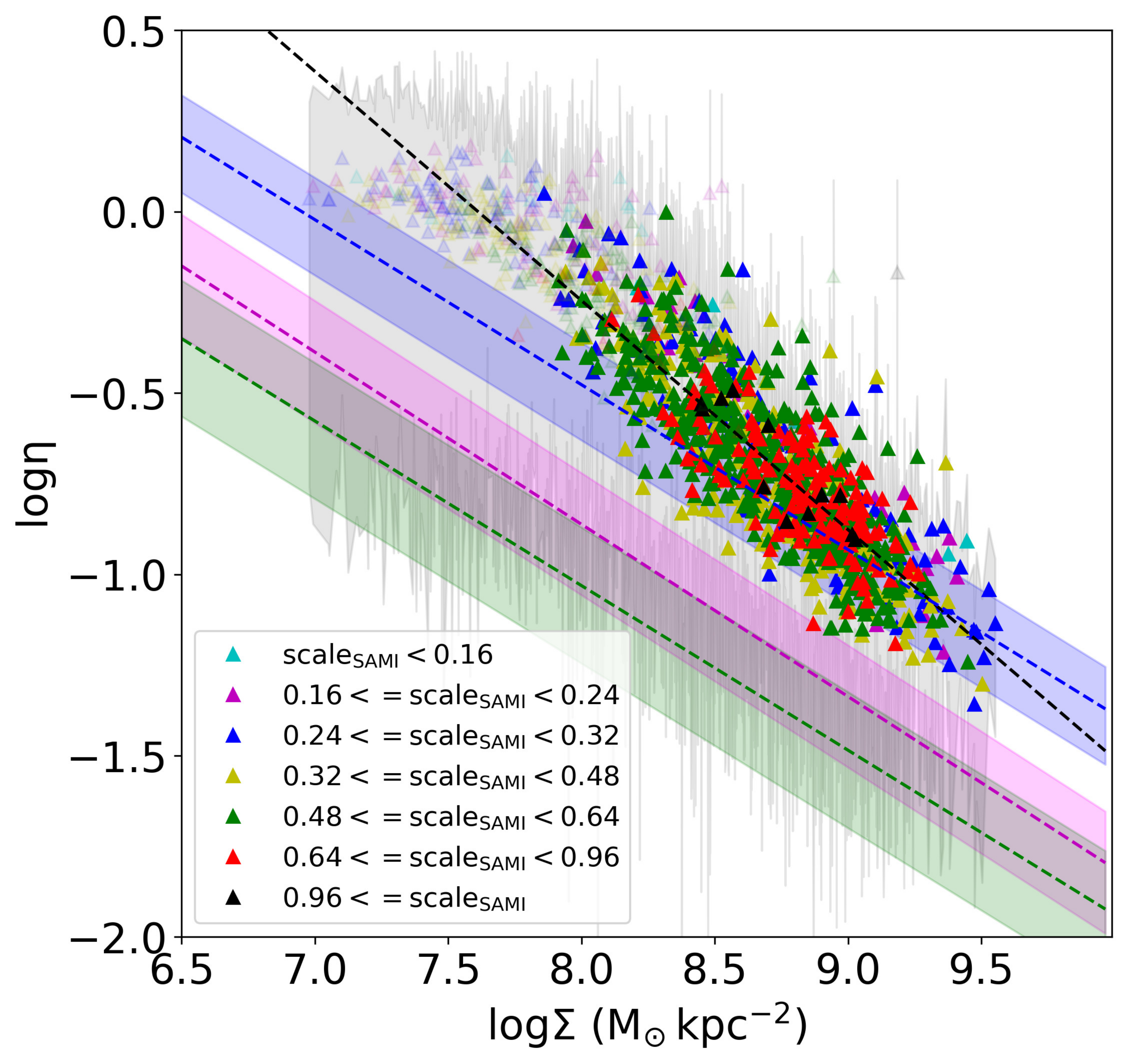}{0.45\textwidth}{}\fig{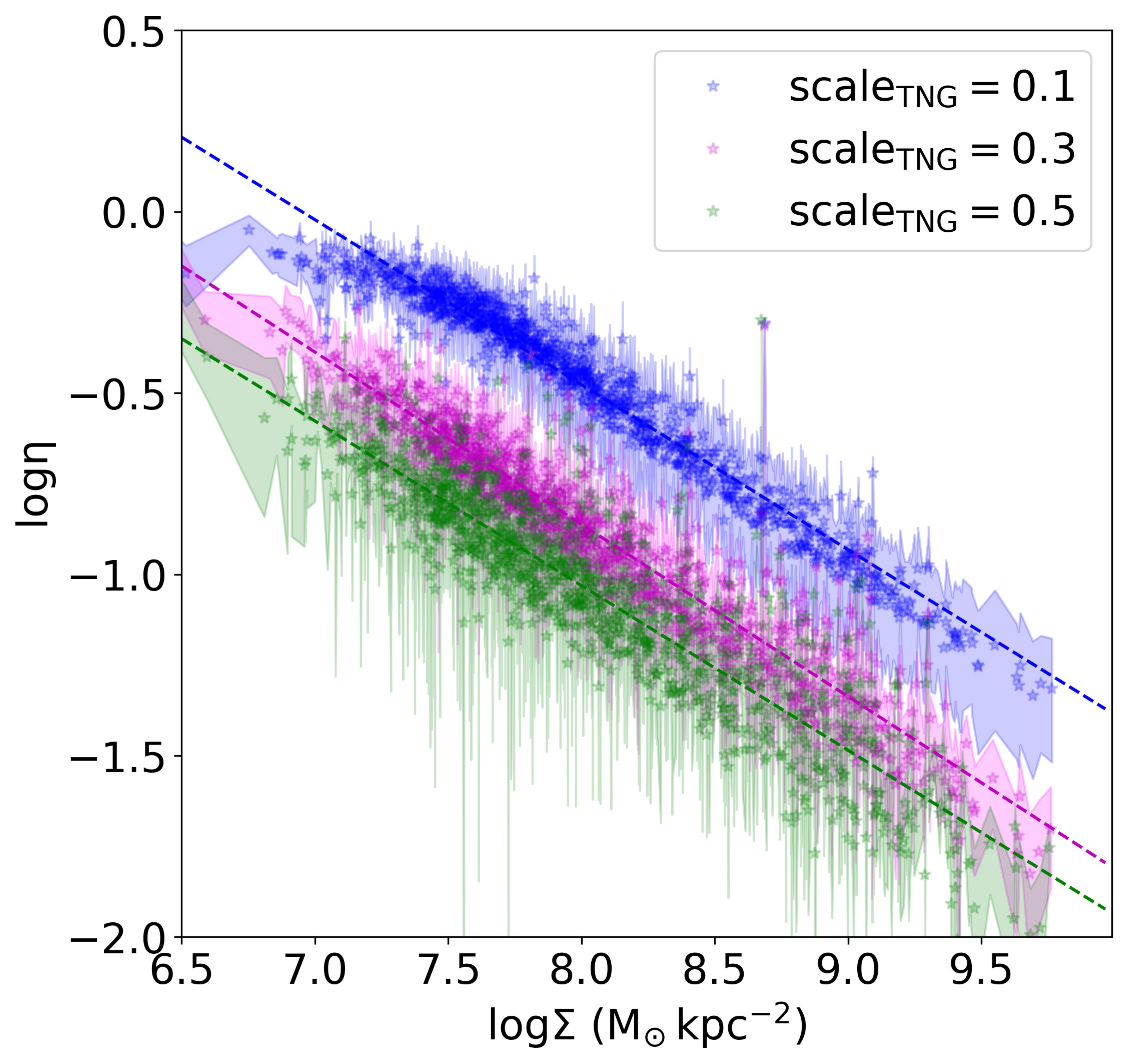}{0.45\textwidth}{}}
	\caption{Comparison of \(\eta\)-\(\Sigma\) Relations between SAMI and {three TNG mock groups}. Left panel: the triangles represent SAMI galaxies, the same as in Figure \ref{fig2}. Dashed lines represent linear fits of SAMI and TNG mock galaxies. {The vertical colored line widths of three mock galaxy groups are the root of the sum of linear fitting standard deviations squared plus average bootstrapping uncertainties squared.} Right panel: the {transparent} pentagrams represent TNG mock galaxies, the different colors represent mock groups of different scales, and color shading of different TNG mock groups shows uncertainties obtained by bootstrapping. The units for all scales are kpc. The slopes of linear fittings of SAMI and TNG are similar, and their intercept differences can be explained by \(V_{\rm rms}\) uncertainties, while intercept differences between the TNG groups are due to their different scales. \label{fig4}
	}
\end{figure*}

Besides observations, we naturally would like to verify them with numerical simulation data. We use TNG50 of the IllustrisTNG project \citep{2019ComAC...6....2N,2019MNRAS.490.3234N,2019MNRAS.490.3196P} to {construct three groups of mock IFU galaxies with mock spaxel scales of 0.1, 0.3, and 0.5 kpc} and calculate their asymmetry parameters. {Similar to the SAMI galaxies, we retain the arbitrary orientation of the mock galaxies {in their natural projections along the principle axes of the simulation box.} Then, the \(180^{\circ}\) rotation is performed relative to the mock galaxy {stellar mass} center. {We also place the center of the mock data cube at the center of each simulated galaxy subhalo.}} The corresponding \(\eta_{\rm\,TNG}\) uncertainties are obtained by the bootstrapping method and are shown with color shading (Figure \ref{fig4}). It should be noted that the contributions of small-scale fluctuations are different at different scales. For example, if the spatial resolution is low, that is, the spaxels of each IFU are large, then fluctuations of smaller scale cannot be resolved (details in Appendix B). We therefore adopt the same scale for each group of mock galaxies to control their small-scale fluctuation degrees. The results show that there are similar relations in the numerical simulation (Figure \ref{fig4}). {We adopted the Emcee \citep{2013PASP..125..306F} Monte Carlo method to linearly fit the pure samples of the SAMI survey and the three TNG mock surveys. Fitting parameters are shown in Table \ref{table1} and dashed lines are shown in Figure \ref{fig4}.} The TNG loglinear relation slopes are similar among different scales, and their intercept differences indicate that their small-scale fluctuations are different. {In principle, due to the degeneracy of measurement uncertainties and small-scale fluctuations, the existence of measurement uncertainties will "lift" the \(\eta\)-\(\Sigma\) relation. \(\eta_{\rm\,SAMI}\) contains both small-scale fluctuations and measurement uncertainties, while \(\eta_{\rm\,TNG}\) contains only the contribution of small-scale fluctuations without measurement uncertainties. {In addition, \(\eta_{\rm\,TNG}\) mostly falls within the maximum uncertainty range indicated by the gray shading of \(\eta_{\rm\,SAMI}\). We thus considered that intercept differences between relations of SAMI and TNG are mainly due to measurement uncertainties of \(\eta_{\rm\,SAMI}\).}} 

\begin{deluxetable}{ccc}
	\tabletypesize{\large}
	\tablenum{1}
	\tablecaption{Loglinear fitting parameters of relations of \(\eta\) and stellar surface mass density \(\Sigma\) \label{table1}}
	\tablewidth{0pt}
	\tablehead{\(\eta=a\Sigma+b\) & \(a\) & \(b\)}
	\startdata
	\(\eta_{\rm\,SAMI}\) & \(-0.632_{-0.039}^{0.040}\) & \(4.811_{-0.352}^{0.346}\)  \\	
	\(\eta_{\rm\,TNG,0.1\,kpc}\) & \(-0.455_{-0.006}^{0.006}\) & \(3.163_{-0.044}^{0.047}\)  \\	
	\(\eta_{\rm\,TNG,0.3\,kpc}\) & \(-0.475_{-0.007}^{0.007}\) & \(2.938_{-0.054}^{0.056}\)  \\	
	\(\eta_{\rm\,TNG,0.5\,kpc}\) & \(-0.454_{-0.007}^{0.007}\) & \(2.601_{-0.053}^{0.052}\)  \\
	\(\eta_{\rm\,TNG,obs}\) & \(-0.578_{-0.008}^{0.008}\) & \(3.697_{-0.064}^{0.065}\)  \\
	\(\eta_{\rm\,TNG,obs,PSF}\) & \(-0.631_{-0.007}^{0.007}\) & \(3.526_{-0.059}^{0.060}\)  \\
	\enddata	
\end{deluxetable}

\begin{figure*}[!ht]
	\gridline{\fig{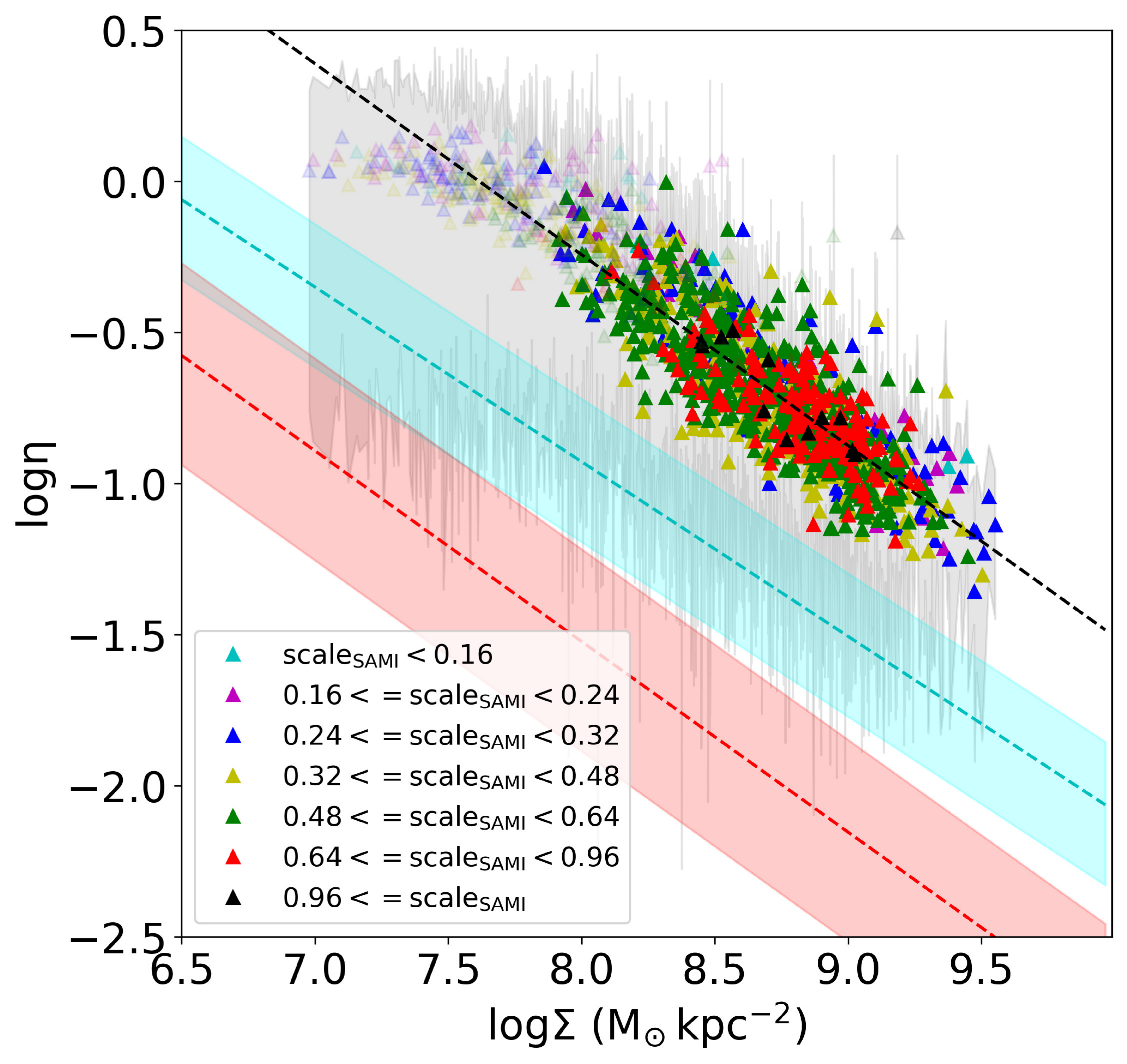}{0.45\textwidth}{}\fig{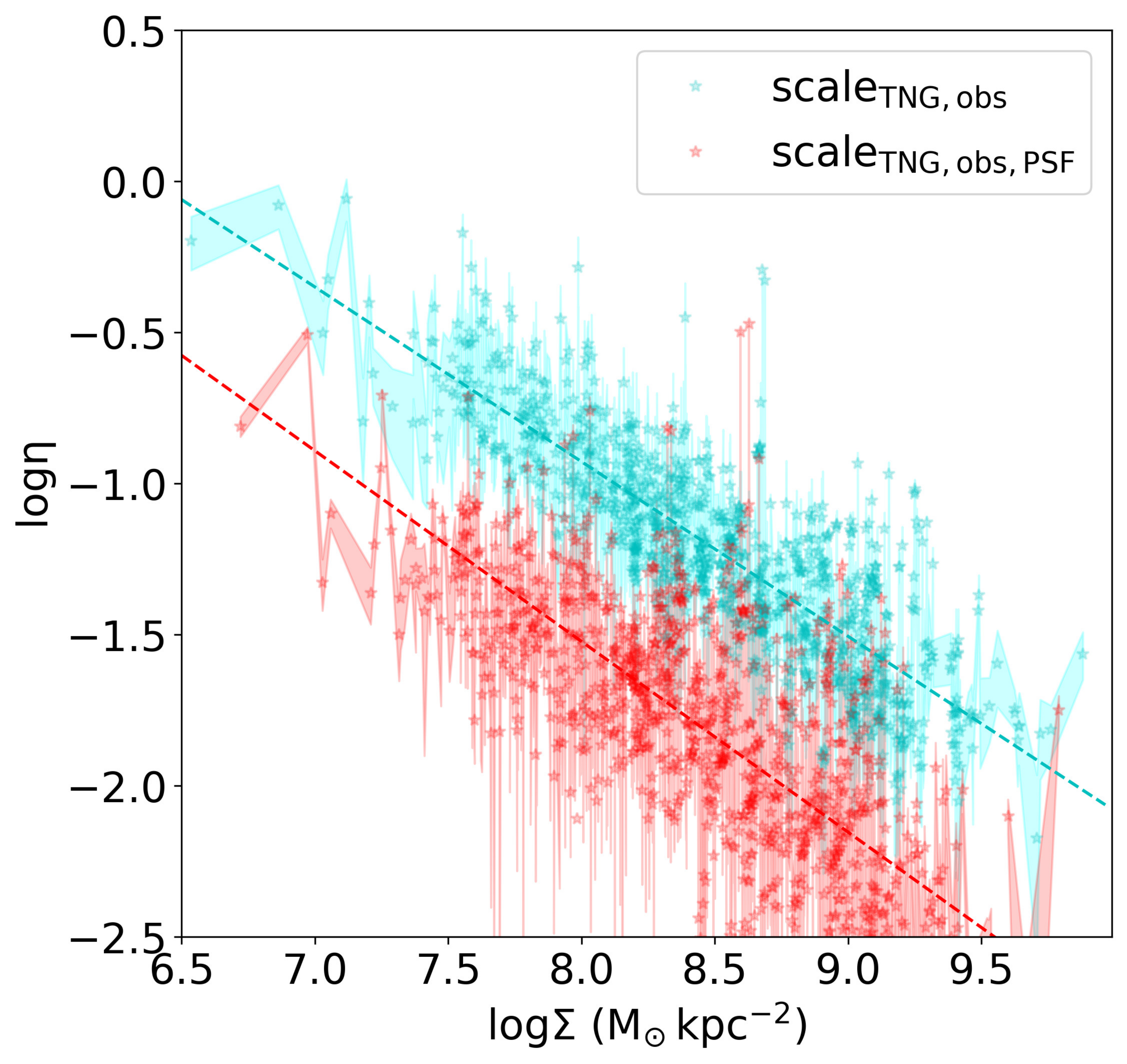}{0.45\textwidth}{}}
	\caption{{Comparison of \(\eta\)-\(\Sigma\) Relations between SAMI and TNG mock observational galaxies. Left panel: the SAMI galaxies, the calculation methods of vertical colored line widths and linear fitting are the same as in Figures \ref{fig2} and \ref{fig4}. Right panel: the TNG mock observational galaxies with or without cosidering PSF influence. We note that the range of y-axis is slightly different from Figure \ref{fig4}. The slopes are more consistent, indicating that the slope difference in Figure \ref{fig4} is mainly caused by the SAMI galaxy being at different distances. The slope is also slightly affected by PSF.} \label{fig4sup}
	}
\end{figure*}

{The \(\eta_{\rm\,TNG}\) can show the essential \(\eta\)-\(\Sigma\) relation of these TNG mock galaxies at three specific distances. However, galaxies in the SAMI sample have different distances. In addition, the observed velocity field is also influenced by the point spread function (PSF). In order to better compare simulation data with observational SAMI galaxies, we constructed a new TNG mock observational sample. For each SAMI galaxy in Figure \ref{fig2}, we selected the TNG galaxy with the closest stellar mass from our TNG sample in Figure \ref{fig4}, and "placed" it at the same distance of the SAMI galaxy, and then we set the size of the mock observational IFU spaxel to \(0^{\prime\prime}.5\), which is the same as all the SAMI IFU spaxels. Therefore, the TNG mock observational galaxy would have the same physical scale (\(\rm kpc\ spaxel^{-1}\)) as the SAMI galaxy, and the corresponding asymmetric parameter is \(\eta_{\rm\,TNG,obs}\). We also investigated the influence of the PSF. We convolved the {stellar mass} and \(V_{\rm rms}\) data cube of the TNG mock observational galaxy with the Gaussian PSF (Appendix A of \citealt{2008MNRAS.390...71C}), while \(\sigma_{\rm PSF}=\rm FWHM_{\rm PSF}/2.355\) and the value of \(\rm FWHM_{\rm PSF}\) are set the same as the original SAMI galaxy of closest stellar mass. The corresponding asymmetric parameter with PSF influence is \(\eta_{\rm\,TNG,obs,PSF}\). The fitting parameters are also shown in Table \ref{table1} and displayed in Figure \ref{fig4sup}. The relation slopes of the two TNG mock observational galaxis are more consistent with that of SAMI. We believe the reason is that, as shown in Figure \ref{fig2}, low-mass galaxies tend to have low surface density. While due to selection effects, they tend to have smaller distances and smaller scales, which is more similar to the upper left corner of the scale 0.1kpc TNG group in Figure \ref{fig4}. On the contrary, high-mass galaxies tend to have larger distances and larger scales, which are more similar to the lower right corner of the scale 0.5kpc TNG group in Figure \ref{fig4}. Therefore, the relation formed by mixing galaxies with different distances will have a steeper slope than the relation of galaxies with the same distances. In another aspect, PSF blur will smooth small-scale pixels and tend to reduce the asymmetric parameter; thus, the \(\eta_{\rm\,TNG,obs,PSF}\) is systematically lower than \(\eta_{\rm\,TNG,obs}\). Moreover, physical scale of PSF influence will be larger on larger-distance galaxies. Due to the selection effect, most of the galaxies in the lower right corner of the relation are farther away from us and will be more severely affected by PSF, thus further increasing the slope value of the overall relaiton. The intercept of the \(\eta_{\rm\,TNG,obs,PSF}\) relation is slightly lower than the uncertainty range of SAMI. There is a possible reason that stellar and dark matter particles of TNG50 has a softening length of 288 pc at redshift \(z=0\) \citep{2019MNRAS.490.3196P}, which may have substantially suppressed the kinematic asymmetry on scales below this. Since the y-axis is in logarithmic, the exact value difference between them is actually not significant. In principle, after adding the measurement uncertainties, the intercept difference of the two relations will be much smaller. We can now conclude that the results of the numerical simulation are in good agreement with the observations. Although the intrinsic \(\eta\)-\(\Sigma\) relation for real galaxies does not necessarily have to be completely consistent with that of TNG, Figures \ref{fig4} and \ref{fig4sup} also indicate that, due to various observational effects, the true intrinsic slope should be less steep than observed.}

\begin{figure}
	\plotone{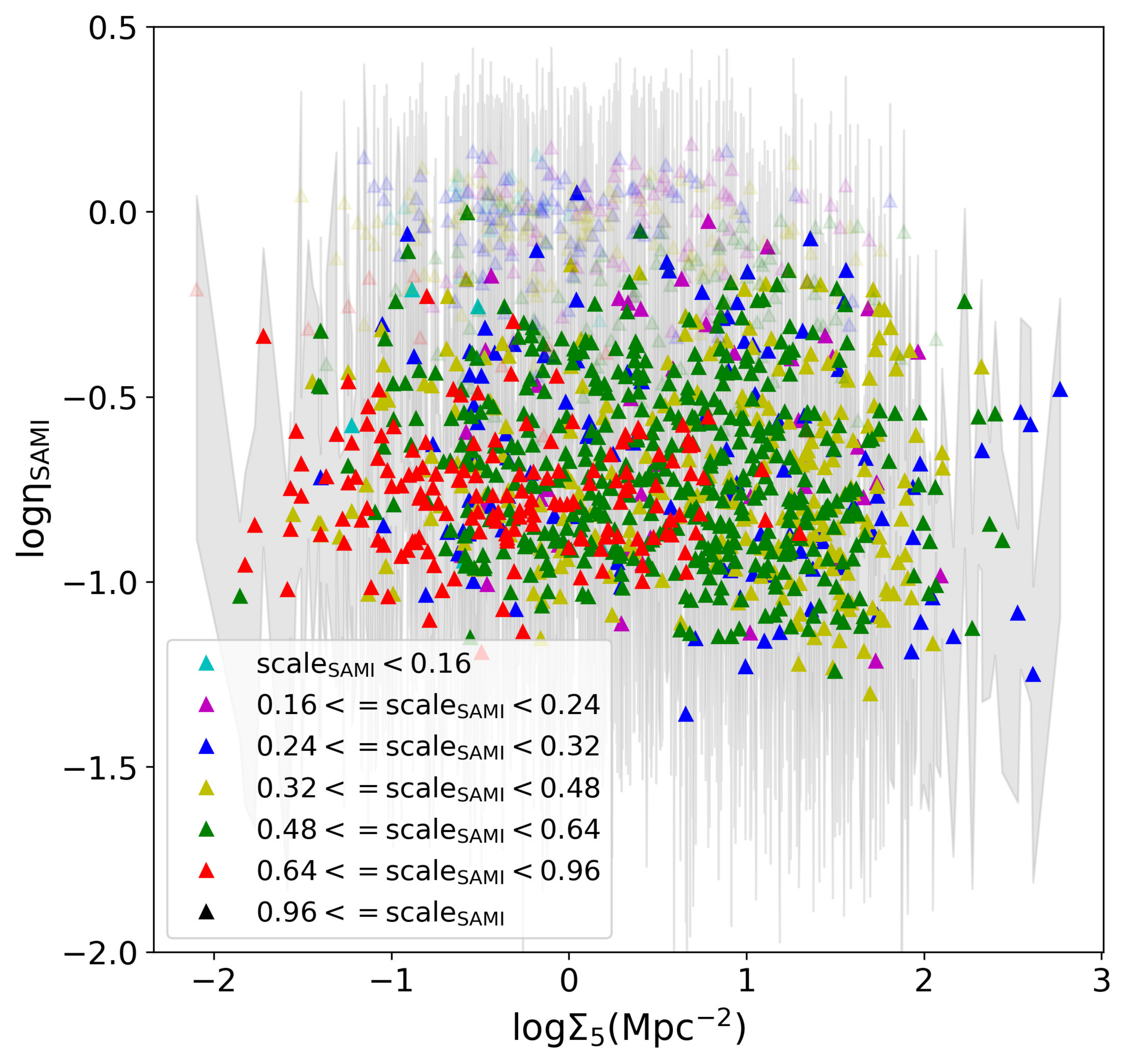}
	\caption{The distribution of \(\eta\) with external environmental metrics \(\Sigma_5\) of SAMI galaxies. There is no obvious correlation between the two parameters, indicating that the small-scale fluctuations are less affected by the external environment of the galaxy and mainly by the internal environment of the galaxy. \label{fig5}
	}
\end{figure}

The kinematical asymmetry of galaxies is usually considered affected by galaxy mergers, interaction, or other influences from external environment \citep{2008ApJ...682..231S,2017MNRAS.465..123B,2020ApJ...892L..20F}. Since the \(\eta\) is related to the stellar surface mass density of the internal environment according to Figure \ref{fig2}, we also investigate its relation with the external environmental factor. We use the SAMI fifth-nearest neighbor surface density \(\Sigma_5\) \citep{2021MNRAS.505..991C}. However, {the fact that \(\eta\) is independent of \(\Sigma_5\) (Figure \ref{fig5}) indicates that small-scale fluctuations are less related to external galaxy environments.} {The gas kinematical asymmetry of SAMI galaxies also exhibits little correlation with fifth or first nearest neighbor distance \citep{2018MNRAS.476.2339B}. Besides, \cite{2018MNRAS.476.2339B} show that gas kinematics is relatively symmetric when high-mass galaxies are not interacting, while low-mass galaxies do not need to be in the interaction stage to have high asymmetry. This result of gas kinematical asymmetry agrees well with our result from Figure \ref{fig2}.} {{In another aspect}, galaxy mergers or interactions usually lead to optical asymmetry, while the inner part of our sample galaxies are optically symmetric according to Figure \ref{fig3}. During sample selection, we also excluded galaxies whose stellar kinematics were affected by mergers or nearby galaxies adopting the catalog from SAMI DR3. All these indicate that the stellar kinematical small-scale fluctuations inside the galaxy shown in Figure \ref{fig2} are not caused by merger, interaction, or other external environment influence.} {Stellar kinematics in galaxies appears to} have its own inherent asymmetry constrained by matter distribution besides external environment influence. Our work also shows that when considering the effect of external environment on kinematical asymmetry, the inherent asymmetry of galaxies themselves should not be ignored.

\begin{figure*}[!ht]
	\gridline{\fig{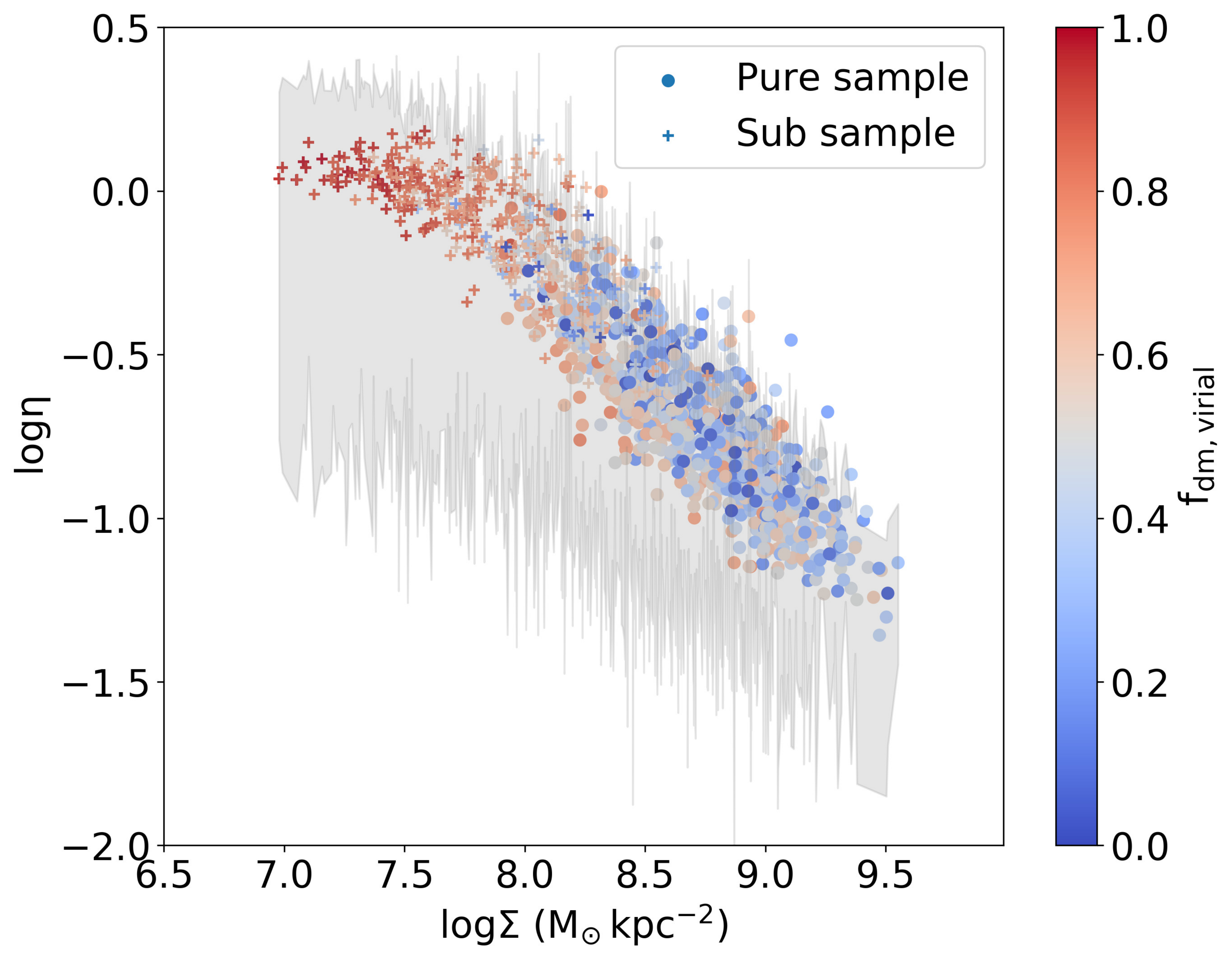}{0.45\textwidth}{}\fig{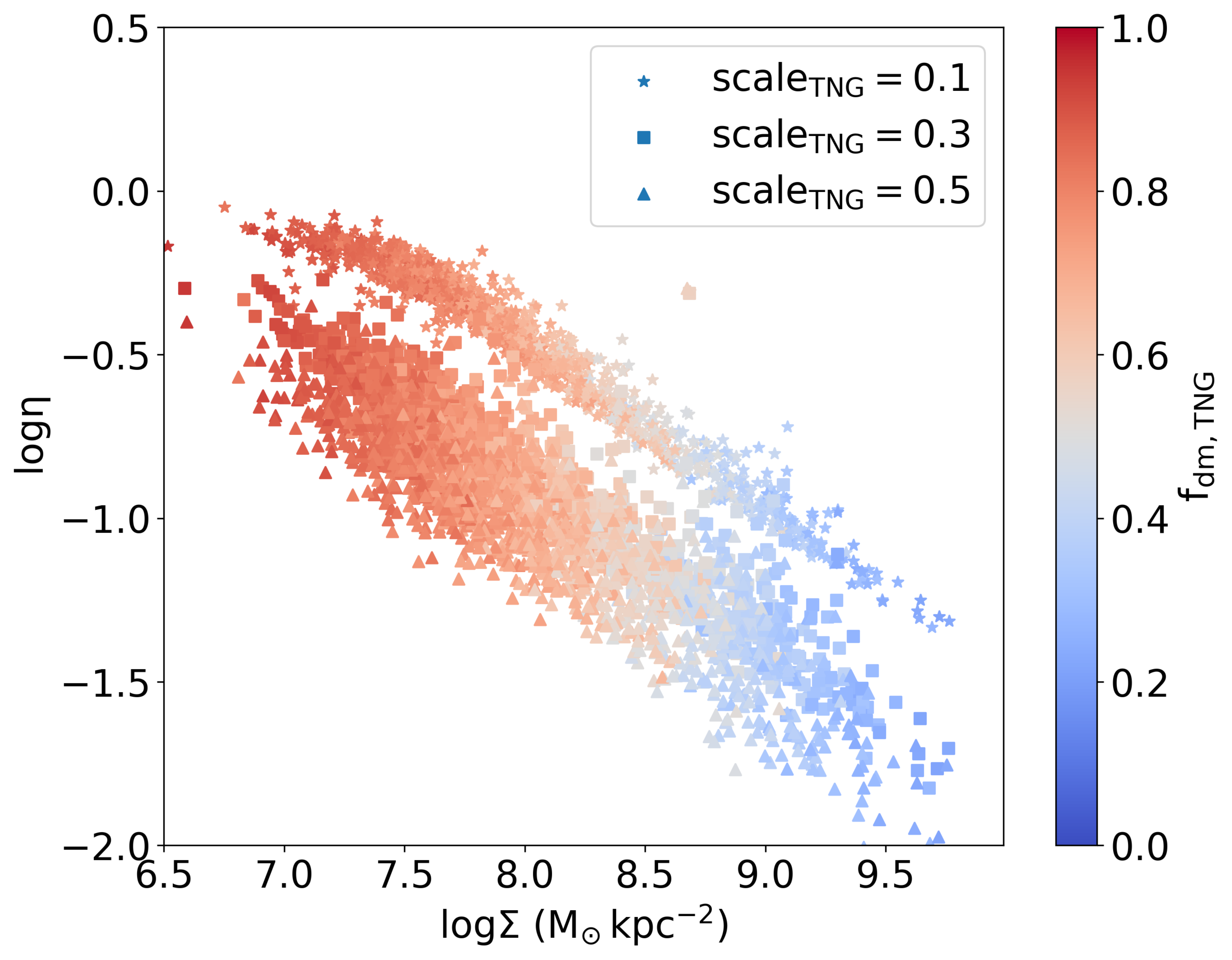}{0.45\textwidth}{}}
	\caption{Dark matter fraction \(f_{\rm dm}\) distribution among the \(\eta\)-\(\Sigma\) relations. Left panel: \(f_{\rm dm,virial}\) is estimated by comparison of galaxy stellar mass with virial dynamical mass. Circles represent the pure samples, and the cross represents the subsamples. Samples with \(f_{\rm dm,virial}\) less than 0 were removed. Shading represents the maximal \(\eta\) uncertainties calculated from \(V_{\rm rms}\) uncertainties. Right panel: \(f_{\rm dm,TNG}\) is obtained by ratio of integration of corresponding mass particles within \(R_{\rm e}\). The color shading of uncertainties is not shown here in order to avoid color confusion, and the units for all scales are kpc. \label{fig6}
	}
\end{figure*}

For galaxies of similar stellar mass to the Milky Way, stars dominate dynamics inside \(R_{\rm e}\). But for low-mass galaxies, especially dwarf galaxies, dark matter {generally dominates} the dynamics inside \(R_{\rm e}\) \citep{2013MNRAS.432.1709C,2019MNRAS.490.2124L}. {{In addition}, the small-scale fluctuations reflect the asymmetry of the stellar kinetic energy per unit mass in the line-of-sight direction, and the kinetic energy distribution is related to the galaxy dynamics. so we considered that theoretically {the small-scale fluctuations can be related to the dark matter mass distribution and dynamics.}} We used dynamical masses obtained from the virial assumption to estimate the galaxy dark matter fraction \(f_{\rm dm}\) \citep{2006MNRAS.366.1126C}, and we also calculate the corresponding values of TNG mock galaxies for comparison (Appendix C). As shown in Figure \ref{fig6}, the dark matter fraction also varies along the relation, that is, galaxy dark matter fractions tend to be higher for the upper left part in the relation and tend to be lower for the lower right part. Although the specific reason for the tight relation of Figure \ref{fig2} is not clear and deserves more research, we {consider} that it should be mainly due to dynamics, such as the dynamical coevolution of stars, gas, and dark matter under gravity.

\section{DISCUSSION}\label{sec4}
It is worth noting again that the measurement uncertainty of the velocity field has a strong influence on the asymmetric parameters. In other words, the contributions of random measurement uncertainties and small-scale fluctuations to the asymmetric parameter are similar and hard to distinguish from each other. we cannot simply use the error propagation formula to calculate the total uncertainty of \(\eta\), which would underestimate the corresponding uncertainty. Therefore, we need to deduct the maximal contribution of \(V_{\rm rms}\) measurement uncertainties and calculate the residual values of \(\eta\), and these residuals can reveal the components that cannot be explained by measurement uncertainties, that is, the minimal contribution of the small-scale fluctuations \(\eta_{\rm subun}\).

The calculation equation is as follows. \(\delta V_{i}^2\) and \(\delta V_{i,\,180}^2\) are corresponding uncertainties obtained by error propagation. The total \(\eta_{\rm subun}\) is also obtained by light-weighted average of \(\eta_{{\rm subun},\,i}\) within \(R_{\rm e}\). 
\begin{equation}
	\eta_{{\rm subun},\,i}={\rm max}(\frac{\left| V_{i}^2-V_{i,\,180}^2\right|-\delta V_{i}^2-\delta V_{i,\,180}^2}{(V_{i}^2+V_{i,\,180}^2)/2},0)
\end{equation}

\begin{figure}
	\plotone{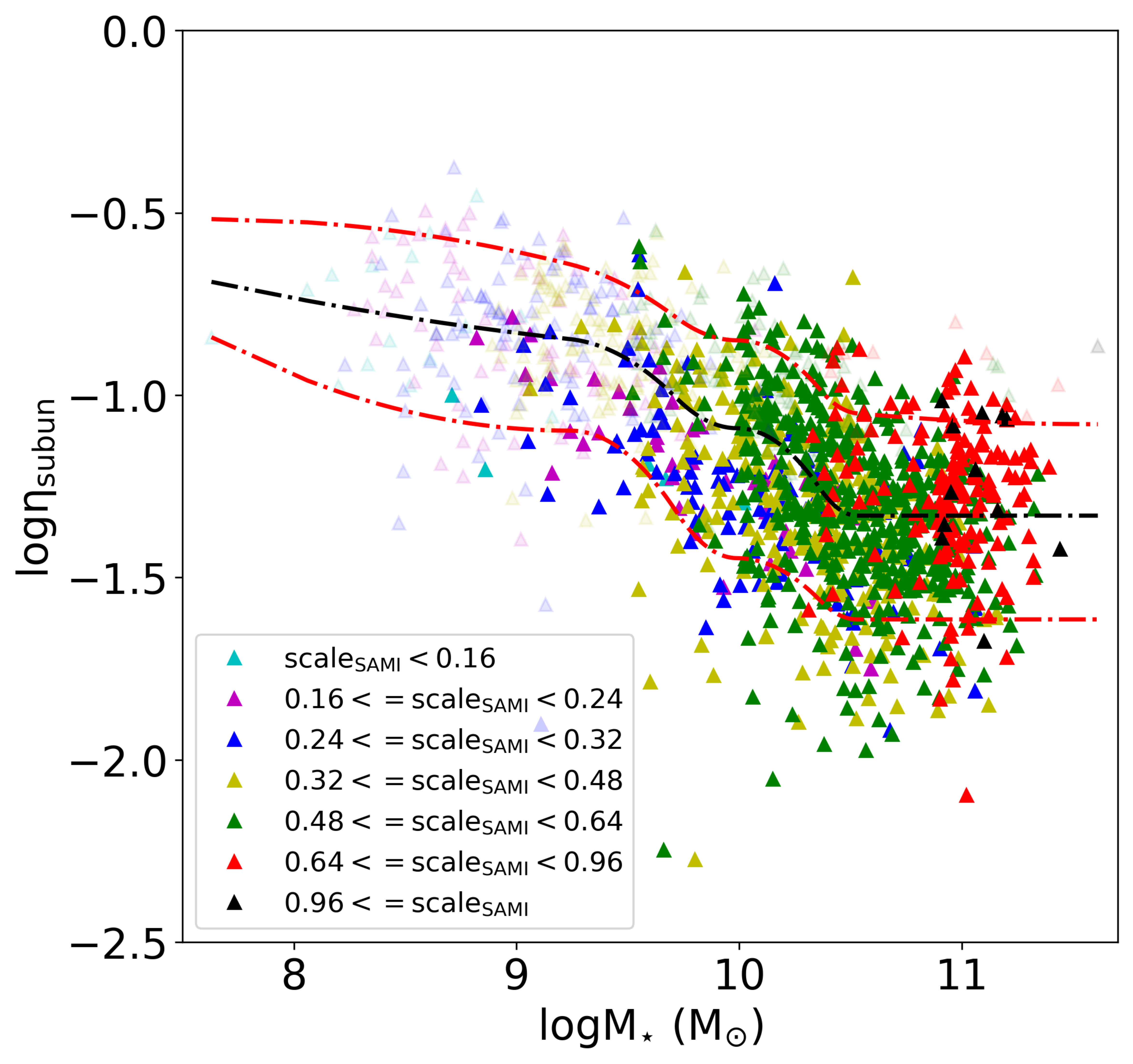}
	\caption{The distribution of \(\eta_{\rm subun}\) after subtracting the maximal contribution of \(V_{\rm rms}\) uncertainties. The relation is similar to the left panel of Figure \ref{fig2}, which indicates that the relation is not a pseudograph caused by the measurement uncertainties of \(V_{\rm rms}\). {The colors of symbols and quantile lines of the samples are the same as those in Figure \ref{fig2}.} \label{fig7}
	}
\end{figure}

We find that the relation persists in the residuals (Figure \ref{fig7}), {and the trend of quantile lines is similar to that in the left panel of Figure \ref{fig2},} indicating that the relation is not a pseudograph caused by the measurement uncertainties of \(V_{\rm rms}\). {It should be noted that the \(\eta_{\rm subun}\) is not the true value of the intrinsic fluctuation degree, since it represents the minimal contribution of the small-scale fluctuations when the uncertainty measurement is reliable, so this \(\eta_{\rm subun}\) is usually lower than the true value.} We take \(\delta\eta=\eta-\eta_{\rm subun}\) to represent the {maximum} uncertainty of \(\eta\){, which is not the standard deviation value of the traditional Gaussian uncertainty but is used to indicate the range where the true value should be}. In most of the figures we use the gray shading to show the corresponding \(\delta\eta\) for each galaxy, and for consistency, the upper and lower shadings take the same \(\delta\eta\). For low-mass galaxies, the measurement uncertainty is relatively large, so {there is a possibility that the uncertainties of their velocity field are} underestimated, then also causing the uncertainty of \(\eta\) to be underestimated, which raises the low-mass side of the left panel of Figure \ref{fig2}, and cannot be reflected by shading length. Nevertheless, nowadays the measurement uncertainty of massive galaxies is reputably accurate. If we only chose massive galaxies of red and black color, we can also obtain the same relation in the right panel of Figure \ref{fig2}, where the relations for galaxies of different colors {appear} to be nearly {the} same. That is to say, even if the measurement uncertainty of the velocity field of low-mass galaxies is somehow underestimated in the SAMI survey, the relation in the right panel of Figure \ref{fig2} should be credible; thus, the small-scale fluctuations that cause the stellar kinematical asymmetry should really exist in galaxies.

{The minimal contribution of the optical asymmetry parameter \(\eta_{\,L\rm subun}\) is similar as follows:
\begin{equation}
	\eta_{{\,L\rm subun},\,i}={\rm max}(\frac{\left| I_{i}-I_{i,\,180}\right|-\delta I_{i}-\delta I_{i,\,180}}{(I_{i}+I_{i,\,180})/2},0)
\end{equation}  	
\(\delta I_{i}\) and \(\delta I_{i,\,180}\) are the corresponding flux uncertainties derived from the SAMI survey. \(\eta_{\,L\rm subun}\) is also obtained by the light-weighted average within \(R_{\rm e}\). We take \(\delta\eta_{\,L}=\eta_{\,L}-\eta_{\,L\rm subun}\) to represent the uncertainty of \(\eta_{\,L}\), which is shown as gray shading in the left panel of Figure \ref{fig3}.
}

\begin{figure}
	\plotone{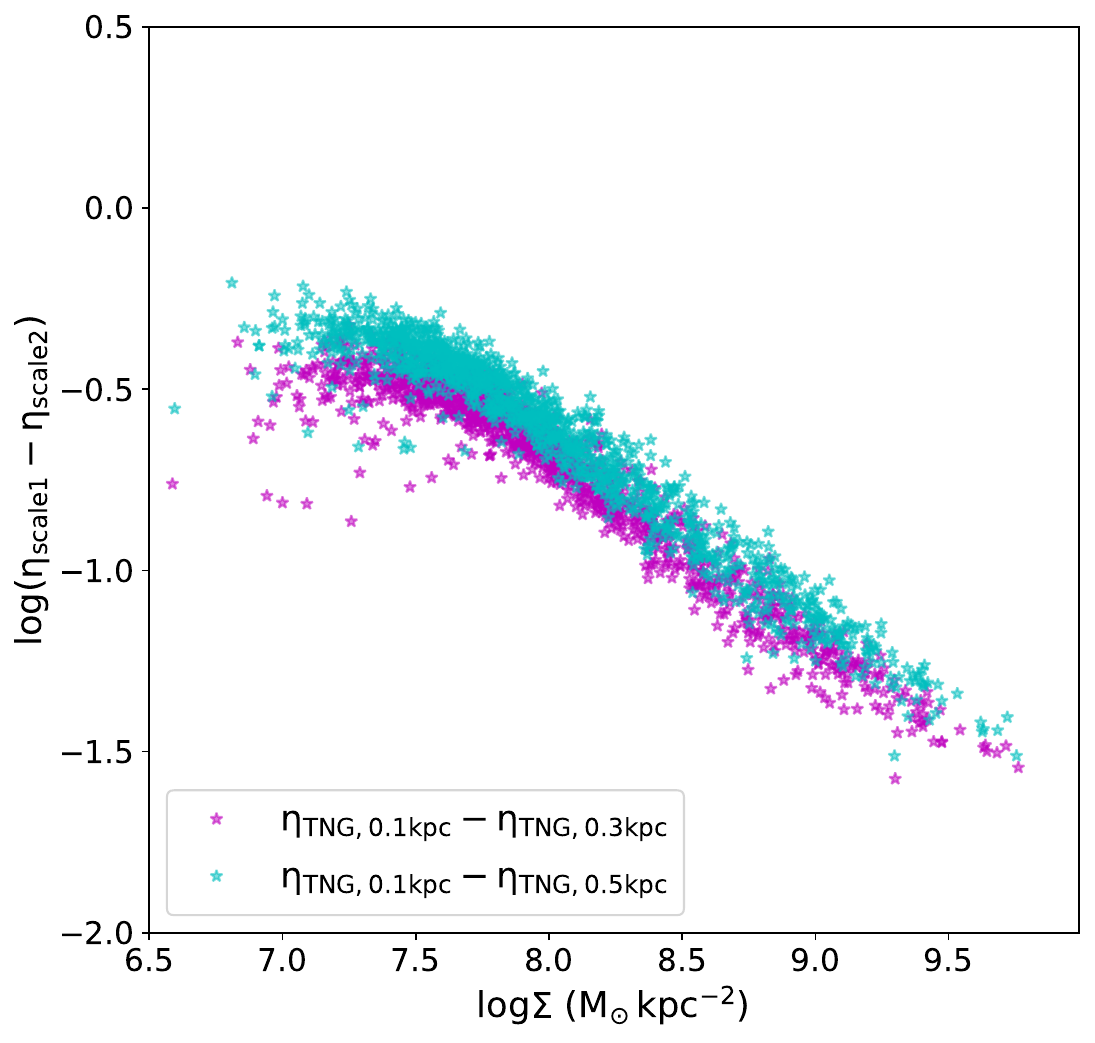}
	\caption{The relation differences between different TNG scale groups. After deducting the contribution of large-scale asymmetry, the relation still exists, indicating that the relation mainly reflects the variation of small-scale fluctuation with stellar mass density. \label{fig8}
	}
\end{figure}

Note that \(\eta\) contains both small-scale and large-scale asymmetries (Appendix B). The small-scale parts are what we call fluctuations here. {Kinematical large-scale asymmetry is closely related to the interaction or merger of galaxies and the resulting optical asymmetry. In another aspect, the small-scale} fluctuation value is affected by the size of the detection scale (IFU spaxel). If the scale {of spaxels} is large, we cannot measure fluctuation values on smaller scales. In principle, the difference in scale only affects the measurement of small-scale fluctuations and does not affect large-scale asymmetries. Therefore, we have used the \(\eta\) subtraction of different scale groups of numerical simulation to deduct the effect of large-scale asymmetry (Figure \ref{fig8}). The residual relation still has similar slopes, indicating that the relation mainly reflects the variation of small-scale fluctuations with mass density. On the other hand, although measuring the specific value of large-scale asymmetry is beyond the content of this article, since the optical distribution of most galaxies is relatively symmetric within \(R_{\rm e}\), {which indicates that they are in relatively stable states. We considered that most galaxies are also large-scale kinematically symmetric within their \(R_{\rm e}\); otherwise, they cannot be in a relatively stable state. That is to say, the contribution of large-scale asymmetry to \(\eta\) is less, and \(\eta\) mainly reflects the amount of small-scale asymmetry.} This is also the reason why asymmetry parameter \(\eta\) can be used to show the relation of small-scale fluctuations.

To sum up, for the first time we found a tight, inverse loglinear relation between stellar kinematical fluctuation degree and galaxy stellar surface mass density. This shows that {although the inner parts of galaxies are generally optically symmetric,} the actual velocity fields of galaxies are not as smooth as the ideal model. There are kinematical small-scale fluctuations inside the galaxy, which cause the stellar kinematics to deviate from symmetry, and the fluctuation degree is mainly related to the internal environment of the galaxy, i.e., the surface density of stars. In addition, the relation also shows that {lower flucations are associated with higher density and lower dark matter fractions, while higher fluctuations are associated with lower density and higher dark matter fractions. The observed relation is also influenced by various observational effects, and the true intrinsic relation slope should be less steep than observed.} The existence of kinematical small-scale fluctuations can have a great influence on the galaxies themselves, which we might have ignored before owing to measurement uncertainty degeneracy. The causes of small-scale fluctuations and the corresponding relation also deserve more research.

\begin{acknowledgements}
We thank X.-X. Xue, D.-D. Xu, L. Zhu, Z. Zheng for early discussion; A. Pillepich for the guide and help with TNG data; and S.-D. Lu for his help with some program codes.

This study is supported by the National Natural Science Foundation of China under grant Nos. 11988101 and 11890694 and the National Key R\&D Program of China No. 2019YFA0405500. Z.-H.Z. acknowledges the scholarship from the International Joint Doctoral Training Program of University of Chinese Academy of Sciences (UCAS).

The data of SAMI survey for this study are available from Australian Astronomical
Optics’ Data Central service (https://datacentral.org.au/). The TNG50 data of the IllustrisTNG project used for this study are available at the website https://www.tng-project.org. The details of the code are presented in Section \ref{sec2}. For reference, the corresponding codes for the quantification of \(\eta\) are available on GitHub at \\ https://github.com/zehaozhong/AsymmetricParameter, or at 10.5281/zenodo.8270377 \citep{zhzh2023}.
\end{acknowledgements}
\software{AsymParaEta \citep{zhzh2023}, astropy \citep{2013A&A...558A..33A,2018AJ....156..123A,2022ApJ...935..167A}, COnstrained B-Spline (cobs) \citep{cobs2007,cobs2022}, Emcee \citep{2013PASP..125..306F}, Matplotlib \citep{2007CSE.....9...90H}, NumPy \citep{2011CSE....13b..22V,2020Natur.585..357H}, Pandas \citep{the_pandas_development_team_2023_8239932}, scipy \citep{2020NatMe..17..261V}}.

\section*{\textbf{Appendix A}}

\textbf{Samples}

\noindent
The input catalogs of SAMI DR3 consist of three parts, three equatorial Galaxy And Mass Assembly (GAMA) regions \citep{2015MNRAS.447.2857B}, eight cluster regions \citep{2017MNRAS.468.1824O} and filter targets. Details of these catalogs are described in  \citet{2021MNRAS.505..991C}. We here only use samples from GAMA regions and cluster regions to study the symmetric parameter, since most filter target galaxies fall outside of the SAMI selection and their kinematical parameters are less reliable or not available. The SAMI Survey team has done many works about both photometry and IFU spectra of galaxies. They adopted multi-Gaussian expansion \citep[MGE,][]{1994A&A...285..723E,2002MNRAS.333..400C} fitting \(r\)-band images from the Sloan Digital Sky Survey or the VST survey and then measured the photometric parameters \citep{2021MNRAS.504.5098D}. They also performed a visual classification of all sample galaxies and measured the galaxy morphologies \citep{2016MNRAS.463..170C}. After cross-matching with their MGE and morphology selection tables, our final sample contained 2100 galaxies from GAMA regions and 896 galaxies from cluster regions. We here only used the stellar kinematics \citep{2017ApJ...835..104V} to calculate the asymmetry parameters.

The IFU images of these galaxies all have the same spaxel size of \(0^{\prime\prime}.5\), but they have different scales (\(\rm kpc\ spaxel^{-1}\)) owing to their different distances. The angular diameter distances are obtained by dimensionless Hubble parameter integration according to the observed spectral redshift of galaxies and are then used to calculate the corresponding arcsec-to-kpc conversion for each galaxy. Scale distribution among our samples is shown in Figure \ref{fig9}. As can be seen, the SAMI samples have observational selection effects owing to the distances of galaxies, so we artificially divided them into seven groups of different scales. However, the contribution of measurement uncertainties to the asymmetric parameter is similar to that of small-scale fluctuation, that is, the two are degenerate. Therefore, the observed relations (Figure \ref{fig2}) cannot distinguish the corresponding scale according to the asymmetric parameters. In addition, We here do not take into account the variance of mass-to-light ratio \(M/L\) inside any galaxy; thus, \(M/L\) is taken as a constant, so we assume that the stellar mass is \(M_{\rm \star}(R_{\rm e})=M_{\rm \star}/2\) within the effective radius \(R_{\rm e}\). Our \(\Sigma\) is also within \(R_{\rm e}\); therefore, we have \(\Sigma=M_{\rm \star}/(2\pi R_{\rm e}^2)\).
\setcounter{figure}{0}
\renewcommand{\thefigure}{A\arabic{figure}}
\begin{figure}
	\plotone{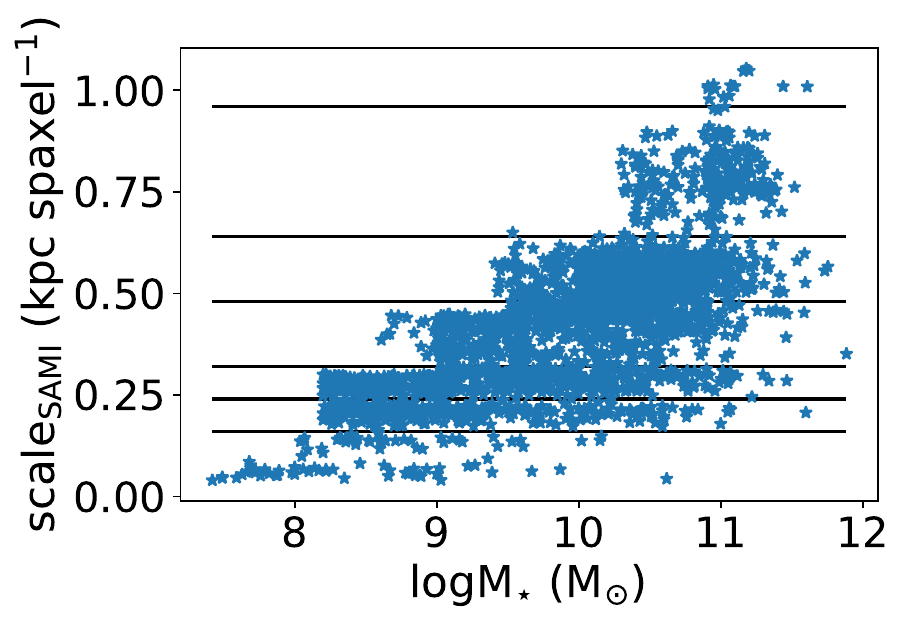}
	
	\caption{The scales per spaxel of SAMI galaxies. We artificially divided the samples into seven groups of different scale regions, which were separated by black lines in the figure. \label{fig9}
	}
\end{figure}

In addition, compared with SAMI DR2, SAMI DR3 added a new measurement of environment density, named fifth nearest neighbor surface density \(\Sigma_5\) \citep{2021MNRAS.505..991C}. Surface density is estimated as \(\Sigma_5=5/(\pi d^2_n)\), where \(d_n\) is the projected comoving distance of the fifth-nearest galaxy within \(\pm1000\ \rm km\ s^{-1}\)  of the SAMI target redshift. We found no obvious correlation between the asymmetry parameter \(\eta\) and \(\Sigma_5\), indicating that the fluctuation of the galaxy is less affected by the external environment and mainly by the internal environment.

TNG \citep{2019ComAC...6....2N} is a suite of the most detailed and widely used galaxy simulations. In order to verify the effect of IFU spaxel scale on the asymmetry parameter \(\eta\), we have used the high mass resolution TNG50 \citep{2019MNRAS.490.3234N,2019MNRAS.490.3196P} to build three groups of mock galaxy surveys, and their IFU spaxel scales are 0.1, 0.3, and 0.5 kpc. Each snapshot of TNG is a set of three-dimensional numerical particles, so we can integrate one dimension and make the rest two-dimensional gridding. For each grid, sum up all the particle masses and weighted average their velocities. Then, each galaxy grid can simulate the observed galaxy IFU spaxel. The original data of these three mock surveys are all the snapshot = 99 group of TNG50-1, that is, redshift \(z=0\). The only difference is that the mock spaxel scales of the two-dimensional grids are different. So each of these three groups of galaxies is a one-to-one correspondence. We use MGE to calculate the \(R_{\rm e}\) of TNG mock galaxies, and then we use Equation (1) to calculate their \(\eta\); the \(\Sigma\) of TNG mock galaxies are also within \(R_{\rm e}\). When we sum up the all \(\eta_{i}\) for every spaxel to the total \(\eta\), there is also a difference between our \(\eta_{\rm\,SAMI}\) and \(\eta_{\rm\,TNG}\). Since the direct observations of the SAMI survey are spectra and the original data of the TNG are mass particles, we use the light-weighted average for the total \(\eta_{\rm\,SAMI}\) and the mass-weighted average for the total \(\eta_{\rm\,TNG}\). The mass resolutions of TNG50 dark matter and stellar particles are both around \(3*10^5\ M_{\rm \odot}\). In principle, the uncertainty of data points at the left low-mass end or the low mass density end (\(\log\Sigma\sim7\ M_{\rm \odot}\ \rm kpc^{-2}\) ) should be greater. We select TNG samples according to the mass range of SAMI samples, so most samples are \(\log(M_{\rm \star}/M_{\rm \odot})\geq8\); thus, they have at least hundreds of particles, so we think the relations should be credible.

We used the bootstrapping method to measure the \(\eta_{\rm\,TNG}\) uncertainties. Each TNG mock galaxy spaxel and line-of-sight orientation can form a square column. The bootstrapping method is to repeatedly randomly select the same number of TNG particles from the sample within the square column, integrate them into two-dimensional spaxels, and then we can form a new bootstrapping mock galaxy. We made 100 bootstrapping mock galaxies for each TNG mock galaxy. The average standard deviation between the corresponding 100 bootstrapping \(\eta_{\rm TNG,boot}\) values and that of the original \(\eta_{\rm\,TNG}\) can be used to measure the uncertainty of the mock galaxy. We have taken these uncertainties to be color shading of the right panel of Figure \ref{fig4}, and the average bootstrapping relative uncertainties of \(\eta_{\rm\,TNG}\) are all around 30\% for the three TNG mock groups. In addition, the average bootstrapping relative uncertainties of \(f_{\rm dm,TNG}\) are all around 0.1\%. We therefore considered our \(\eta_{\rm\,TNG}\) and \(f_{\rm dm,TNG}\) values to be reliable.

For the cosmological parameters, we use the same values as for TNG50. That is, the dimensionless Hubble constant \(h=0.6774\), the total matter density \(\Omega_m=0.3089\) and the dark energy density \(\Omega_{\Lambda}=0.6911\). 
\hspace*{\fill} \\
\setcounter{table}{0}
\renewcommand{\thetable}{B\arabic{table}}
\begin{table*}
	%\tabletypesize{\scriptsize}
	\large
	%\tablenum{2}
	\caption{Schematic diagrams of large-scale asymmetry and small-scale fluctuation \label{table2}} 
	\begin{center}
		\begin{tabular}{|c|c|c|c|}
			\hline
			\(20^2\) & \(20^2\) & \(21^2\) & \(21^2\) \\
			\hline
			\(20^2\) & \(20^2\) & \(21^2\) & \(21^2\) \\
			\hline
		\end{tabular}
		\begin{tabular}{|c|c|c|c|}
			\hline
			\(20^2\) & \(20^2\) & \(200\) & \(20^2\) \\
			\hline
			\(20^2\) & \(20^2\) & \(20^2\) & \(600\) \\
			\hline
		\end{tabular}
	\end{center}
\end{table*}

\section*{\textbf{Appendix B}}
\textbf{Significant effect of spaxel scales on \(\eta\)}

\noindent
The scale size of the IFU spaxel has a significant effect on \(\eta\), because \(\eta\) essentially consists of contributions from both large scales and small scales, which we refer to here as "large-scale asymmetry" and "small-scale fluctuation," respectively. A schematic diagram is shown in Table \ref{table2}. Assume the two galaxies in Table \ref{table2} contain 8 spaxels and each spaxel has the same weight. The unit of each spaxel is \((\rm km\ s^{-1})^2\), and the center of the table is the galaxy center. Then, the left side of the table shows the large-scale asymmetry, and the right side of the table shows the small-scale fluctuation. Their asymmetry parameters are \(\eta=0.0975\) and \(\eta=0.2667\), respectively, according to Equation (1). If the scale is doubled, then the two galaxies remain 2 spaxels. The galaxy on the left side of Table \ref{table2} becomes \([20^2,21^2]\), which is still asymmetric. However, the galaxy on the right side of Table \ref{table2} becomes \([20^2,20^2]\), which be symmetric pattern after weighted averaging. From this example we can obtain \(\eta_{\rm doubled}\leq\eta\), and generally there is roughly \(\eta_{\rm larger-scale}\leq\eta_{\rm smaller-scale}\) for the same galaxy. This schematic diagram shows that if the IFU spaxels are too large, contributions of smaller-scale fluctuations would be missed. The results for the three mock groups of TNG in the right panel of Figure \ref{fig4} can also indicate that the larger the scale, the smaller the calculated asymmetry parameter \(\eta\) for the same galaxy.
So the differences in intercepts between the three groups represents their different contributions to \(\eta\).

\section*{\textbf{Appendix C}}
\textbf{The calculation of dark matter fraction \(f_{\rm dm}\)}

\noindent
It should be noted that in this work the "dark matter fraction" within \(R_{\rm e}\) actually contains the gas mass. But most early-type galaxies have little gas, and for those other galaxies or spiral galaxies that contain a lot of gas, there is usually little gas within \(R_{\rm e}\). In addition, it is difficult to measure total gas mass accurately for IFS observations. Therefore, we here mainly focus on the "dark matter fraction" variation along the \(\eta\)-\(\Sigma\) relation and statistically ignore the gas contributions.

The definition of dark matter fraction here is \(f_{\rm dm}=1-M_{\rm \star}/M_{\rm dyn}\), where \(M_{\rm dyn}\) means dynamical mass, and thus the total mass of galaxies. For all samples of SAMI galaxies, we calculate their virial dark matter fraction \(f_{\rm dm,virial}\), as shown in Figure \ref{fig6}. Since it was the virial estimated value, we removed a few samples of \(M_{\rm \star}>M_{\rm dyn}\) (thus \(f_{\rm dm}<0\)). The virial dynamical masses are \(M_{\rm dyn}=\beta R_{\rm e}\sigma_e^2/G\), and the best factor we used is \(\beta=5.0\)  \citep{2006MNRAS.366.1126C}. So the virial \(M_{\rm dyn}\) we obtained here and the corresponding virial \(f_{\rm dm}\) are only approximate values. Figure \ref{fig6} can preliminarily reflect the correlation between \(f_{\rm dm}\) and the \(\eta\)-\(\Sigma\) relation, and more precise values need to be calculated for future work.

For \(f_{\rm dm,TNG}\), we calculate them of within \(R_{\rm e}\), and we have \(f_{\rm dm,TNG}=1-M_{\rm \star,TNG}(R_{\rm e})/M_{\rm all,TNG}(R_{\rm e})\). The \(M_{\rm \star,TNG}(R_{\rm e})\) and \(M_{\rm all,TNG}(R_{\rm e})\) can be obtained by summing the masses of the corresponding particles in the three-dimensional \(R_{\rm e}\) range to the galaxy centers.
\\
%\addtocounter{figure}{-4}
%\captionsetup[figure]{labelfont=bf, name=Figure, labelsep=period}
%\renewcommand{\thefigure}{S\arabic{figure}}

%\captionsetup[table]{position=above} 
%\captionsetup[table]{labelfont=bf, labelsep=period}
%\renewcommand{\tablename}{Extended Data Table}
%\renewcommand{\thetable}{S\arabic{table}}
%\renewcommand{\arraystretch}{2} 

\turnoffediting
\bibliography{kinesym06_bib}
\bibliographystyle{aasjournal}
%\renewcommand\refname{REFERENCES AND NOTES}
%\bibliography{kinesym06_bib}
%\bibliographystyle{Science}
\end{document}